\documentclass[12pt]{article}

\usepackage{amsmath,amssymb,amscd}
\usepackage{caption}

\usepackage{graphicx}

\newcommand{\beq}{\begin{equation}}
\newcommand{\eeq}{\end{equation}}
\newcommand{\nbea}{\begin{align*}}
\newcommand{\neea}{\end{align*}}
\newcommand{\nbeq}{\begin{equation*}}
\newcommand{\neeq}{\end{equation*}}

\newcommand{\ETslash}{\ensuremath{/ \hspace{-.7em} E_T}}

\usepackage{array}
\newcolumntype{M}[1]{>{\centering\arraybackslash}m{#1}}
\newcolumntype{N}{@{}m{0pt}@{}}

\numberwithin{equation}{section}

\begin{document}

\begin{titlepage}

\pagestyle{empty}

\baselineskip=21pt
\rightline{\footnotesize KCL-PH-TH/2018-32, CERN-TH/2018-153}
\vskip 0.75in

\begin{center}

{\large {\bf Phenomenological Constraints on Anomaly-Free Dark Matter Models}}

\vskip 0.5in

 {\bf John~Ellis}$^{1,2,3}$,~
   {\bf Malcolm~Fairbairn}$^{1}$
and {\bf Patrick~Tunney}$^{1}$

\vskip 0.5in

{\small {\it

$^1${Theoretical Particle Physics and Cosmology Group, Physics Department, \\
King's College London, London WC2R 2LS, UK}\\
\vspace{0.25cm}
$^2${NICPB, R\"avala pst.~10, 10143 Tallinn, Estonia}\\
$^3${Theoretical Physics Department, CERN, CH-1211 Geneva 23, Switzerland}
}}

\vskip 0.5in

{\bf Abstract}

\end{center}

\baselineskip=18pt \noindent


{\small
We study minimal benchmark models of dark matter with an extra anomaly-free U(1)$^\prime$ gauge boson Z$^\prime$.  
We find model parameters that give rise to the correct cosmological dark matter density while evading the 
latest direct detection searches for dark matter scattering produced by the XENON1T
experiment, including the effects of $Z-Z'$ mixing.  We also find regions of parameter space that evade
the constraints from LHC measurements of dileptons and dijets,
precision electroweak measurements, and LHC searches for monojet events with missing transverse energy, 
$\ETslash$. We study two benchmark Z$^\prime$ models with $Y$-sequential couplings to quarks
and leptons, one with a vector-like coupling to the dark matter particle and one with an axial dark matter coupling.
The vector-like model is extremely tightly constrained, with only a narrow allowed strip where $m_\chi \simeq M_{Z'}/2$,
and the axial model is excluded within the parameter range studied. We also consider two leptophobic Z$^\prime$ benchmark models, finding
again narrow allowed strips where $m_\chi \simeq M_{Z'}/2$ as well as more extended regions where $\log_{10} (m_\chi/ {\rm GeV}) \gtrsim 3.2$.}


\vskip 0.75in

\leftline{ {
July 2018}}

\end{titlepage}

\section{Introduction}

The existence of the dark matter required by astrophysics and cosmology \cite{Zwicky,Rubin, Peebles,planck} is one of the most pressing arguments for physics beyond the Standard Model, and its nature remains a mystery, despite many theoretical proposals and experimental searches. The simplest explanation is that the dark matter is some species of massive particle, and if this interpretation is correct the dark matter should be provided by some particle beyond the Standard Model. However, the range of possible dark matter particle masses is very broad, extending from the Planck mass down to $\ll$~eV. Within this range, one of the favoured possibilities is some type of
weakly-interacting massive particle (WIMP) that was in thermal equilibrium with Standard Model particles during the early history of the Universe, but decoupled as it expanded and cooled. The typical range of WIMP masses that give rise to a good relic density today is in the GeV to TeV range, placing these particles potentially within reach of experiments at the LHC as well as direct and indirect searches for astrophysical dark matter. The prototypical WIMP candidate was a massive
sequential neutrino~\cite{leeweinberg,Hut,hutolive}, but this has been ruled out by a 
combination of accelerator (see, for example, \cite{LHC4thneutrino}) and non-accelerator experiments.
Many WIMP candidates from scenarios for physics beyond the Standard Model have been proposed subsequently~\cite{BertoneHooperSilk}, one
of the most prominent being supersymmetry~\cite{EHNOS}. This theory has many potential experimental signatures beyond the WIMP
particle itself, but also has many free parameters. Thus, although no experiment has found any evidence for supersymmetry,
its appearance at the TeV scale cannot yet be ruled out. That said, interest has developed in exploring alternative
WIMP scenarios. 

In the absence of clear theoretical guidance, much activity has gone into the formulation and testing of simplified
dark matter models that involve only a small number of relevant parameters, which can in principle be explored
systematically. These simplified dark matter models may be divided into categories according to the way the dark matter candidates
interact with Standard Model particles. The focus has evolved from effective field theories of these dark matter
interactions~\cite{eft1,eft2,moreoneft} to more complete dynamical models featuring mediator particles, usually bosons of spin zero or one~\cite{busoni,mccabe, WhitePapers}.
In principle, the mediator particle could be the Higgs or $Z$ boson of the Standard Model, scenarios that are
tightly constrained, but not excluded~\cite{zportal,higgs1,higgs2,Estonia}.

Here we consider the alternative scenario in which the mediator is a boson that is not included in the
Standard Model. These mediator particles could be produced at the LHC as well as the dark matter particles themselves,
and the masses and couplings of the mediator particles are also constrained by the cosmological dark matter density,
as well as by direct and indirect searches for astrophysical dark matter. We study here the possibility of a single mediator particle $Z'$ with spin one.  Extensions of the standard model containing a new $Z'$  are extremely well studied in the literature going back several decades \cite{langacker1984,erler,appelquist,Carena,morrissey,chiang,langacker}. Such models feature the possibility of mixing with the $Z$ boson, which is
constrained by precision electroweak measurements. Moreover, they are strongly constrained by gauge invariance. In particular, the ultraviolet
completions of these models should be free of triangle anomalies \cite{barr,batra,ekstedt,Ismail}.

A complete `simplified' model of dark matter should include some mechanism for cancelling these triangle anomalies,
which could in principle be achieved in different ways \cite{Kahlhoefer:2015bea,ekstedt}. The option we pursue in this paper is
that the anomalies are cancelled by new physics at the TeV scale, which entails an interesting new set of phenomenological 
signatures and possible experimental constraints~\footnote{The alternative is to assume that the anomaly-cancellation
mechanism operates at some high energy scale, generating anomalous, apparently non-renormalizable gauge-boson interactions that are also detectable 
in principle at lower energies~\cite{Katz}.}.
Since there are, in total, six different gauge anomalies to be cancelled, the constraints on the beyond the
Standard Model fermions needed to cancel them are non-trivial~\cite{usEFT1}. Consequently, the minimal `simplified' dark matter
models cannot always be as simple as those originally considered, and the phenomenological signatures are correspondingly
more complex and interesting~\footnote{For other studies of anomaly-free $Z'$ models in the context of dark matter, see~\cite{Duerr,Ismail}.}.

In a previous paper~\cite{usEFT1} we constructed systematically specific minimal anomaly-free dark matter models
with a U(1)$^\prime$ boson $Z^\prime$ whose couplings to quarks and leptons are generation-independent~\footnote{See~\cite{usEFT2}
for the generalization to anomaly-free $Z'$ models motivated by deviations from the Standard Model in
$B \to K^{(*)} \ell^+ \ell^-$ decays.}. The simplest such models are leptophilic, and
are subject to various powerful experimental constraints. In particular, the LHC constraint on resonances in
dilepton mass spectra is now very strong, imposing important restrictions on U(1)$^\prime$ models in which the $Z^\prime$
boson couples to the charged leptons $e^+ e^-$ and $\mu^+ \mu^-$~\cite{tytgat}. Another powerful constraint comes from
direct searches for dark matter scattering on nuclei, in which the market leader is now the XENON1T experiment~\cite{XENON1T}.
This constraint is particularly important for U(1)$^\prime$ models in which the $Z^\prime$
boson has vector-like couplings to Standard Model particles and/or dark matter, since coherent enhancement leads to an enhanced cross section in these situations. These considerations motivate specific studies of benchmark U(1)$^\prime$ models in which the $Z^\prime$ boson is either leptophobic and/or has axial couplings, as also discussed in~\cite{usEFT1}.

We found in~\cite{usEFT1} that models with a single dark matter particle necessarily contain a leptophilic $Z^\prime$ with couplings
to quarks and leptons that are proportional to those in the Standard Model - such models have become known as $Y$-sequential models \cite{appelquist,ekstedt}. 
In such models, $Z' - Z$ mixing is unavoidable, inducing important contributions to precision electroweak observables that impose a powerful
 constraint on $M_{Z'}$ \cite{sandt2018}. Moreover, the dark matter particle must have vector-like $Z'$ couplings. Because
of these two features, the experimental constraints on this benchmark model are very strong, 
as we discuss in detail in Section~\ref{sec:benchmark} of this paper, and only a very small region of the model's parameter space survives.

In Section~\ref{sec:axial} we then discuss a second $Y$-sequential benchmark model in which the dark matter particle has axial $Z'$ couplings, 
with the aim of reducing the impact of the direct dark matter
search experiments. However, the dark matter density constraint is more important in this case,
the $Z^\prime$ is still leptophilic, and there is again an important constraint from precision electroweak data. 
Thus, even though the direct dark matter scattering constraint
has less impact, the other constraints are still sufficiently powerful to exclude this model within the parameter
range we explore~\footnote{As discussed in~\cite{usEFT1},
this axial dark matter particle must be accompanied by at least one other `dark' particle with a
U(1)$^\prime$ charge, whose production offers in principle a distinctive $\ETslash$ signature at the LHC.
However, we do not discuss it in this paper.}.

Therefore, in Section~\ref{sec:leptophobic} we also consider making the $Z^\prime$ leptophobic, which requires at least 
two additional particles in the dark sector, with non-zero Standard Model charges. We consider two benchmark scenarios
proposed in~\cite{usEFT1}, one with SU(2) doublet dark sector particles in which the $Z'$ couplings to quarks are suppressed, and LHC monojet constraints
become important, and another with SU(2) triplet dark sector particles in which the quark couplings are less suppressed,
so that the LHC dijet constraints are more important. In both cases the direct dark matter search constraint is more restrictive,
but allows extended regions where $\log_{10} (m_\chi/ {\rm GeV}) \gtrsim 3.2$.

Finally, we present our conclusions and some discussion in Section~\ref{sec:conclusions}.

\section{Benchmark with a Single Dark Matter Particle \label{sec:benchmark}}

We consider first the possibility that the only dark sector particles are fermions that are uncharged singlets of the Standard Model
gauge group. Restricting our attention to generation-independent U(1)$^\prime$ charge assignments,
denoting the left-handed lepton doublets by $l$, the right-handed lepton singlets by $e$, 
the right-handed quark singlets by $u, d$ and the left-handed quark doublets by $q$, and choosing the normalisation $Y'_q = 1$,
we found~\cite{usEFT1} the following unique solution:
\begin{equation}
Y'_l \; = \; -3, \quad Y'_e \; = \; -6, \quad Y'_d \; = \; -2 , \quad Y'_u \; = \; 4 ,  \quad Y'_H \; = \; -3\, ,
\label{Bench1}
\end{equation}
which is known in the literature as the $Y'$-sequential model~\cite{Yseq,ekstedt}. Its free parameters include the U(1)$^\prime$ gauge coupling $g$
and the masses of the $Z^\prime$ and the dark matter particle $\chi$. 
If there is a single particle in the dark sector, it must be vector-like under U(1)$^\prime$: $Y'_{\chi,L} = Y'_{\chi,R}$~\cite{usEFT1},
but the magnitude of the U(1)$^\prime$ charge of this dark matter particle is arbitrary, introducing a fourth parameter into this minimal model. 

\begin{figure}
\centering
\includegraphics[width = 0.6\textwidth]{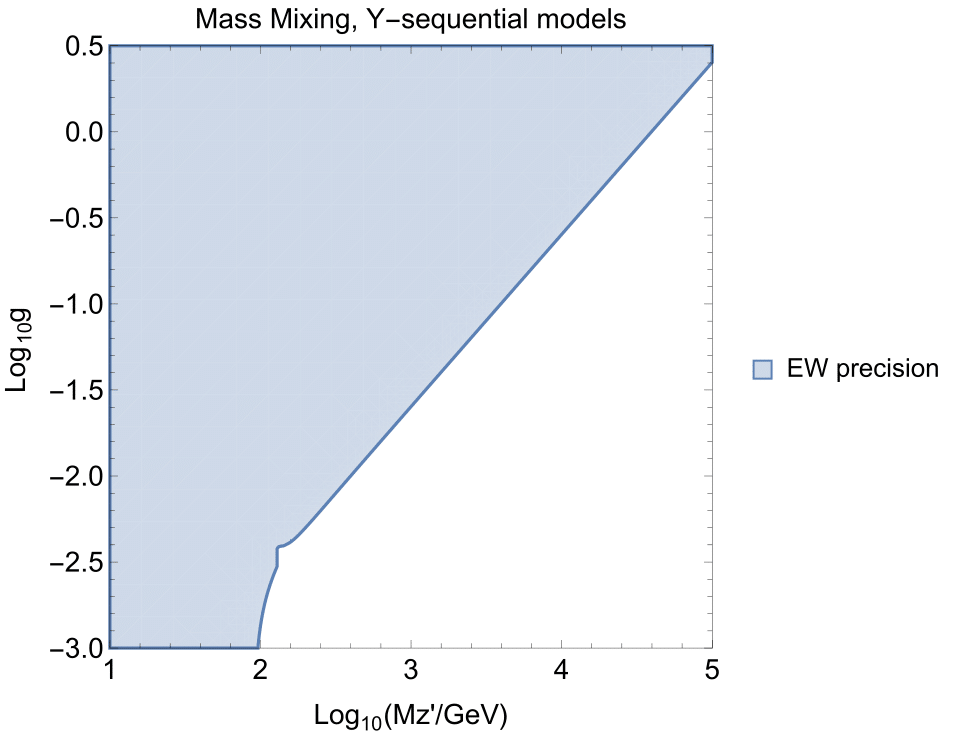}
\caption{\it The $(M_{Z^\prime}, g)$ plane in the U(1)$^\prime$ Y-sequential model, showing the impact of the
constraints on the oblique parameters $S$ and $T$ imposed by precision electroweak measurements.}
\label{fig:EWPrecision}
\end{figure}

We consider next the constraints on the Y-sequential model that are imposed by precision electroweak
measurements, specifically the constraints from the oblique parameters $S$ and $T$. 
As seen in Eq.~(\ref{Bench1}), this and other $Y$-sequential models have the feature that the Higgs doublet has a non-zero
U(1)$^\prime$ charge. Consequently, tree-level $Z^\prime - Z$ mixing is unavoidable, and is calculable as a function of the
U(1)$^\prime$ gauge coupling $g$ and the $Z^\prime$ mass, increasing as $g$ increases and/or 
$M_{Z^\prime} \to M_Z$.   Therefore the precision electroweak constraint is stronger in these cases, as seen in Fig.~\ref{fig:EWPrecision}~\footnote{In 
this Section we neglect kinetic mixing, since mass mixing is much much more important in these leptophobic models. The details of mass and kinetic
mixing are described further in Appendix~\ref{AppMix}.}.
This mixing also has important implications for 
the calculations of the relic density of the dark matter particle,
$\Omega_\chi h^2$ and of the dark matter scattering cross section, which we discuss below.

If the dark sector contains more than one particle, it is possible
that $Y'_{\chi,L} \ne Y'_{\chi,R}$. As already advertised, in order to minimise the impact of
direct dark matter searches, the case where the dark matter particle has a purely axial $Z'$ coupling,
$Y'_{\chi,L} = - Y'_{\chi,R}$, is of particular interest. The electroweak precision constraint shown in Fig.~\ref{fig:EWPrecision}
is applicable to that model as well as to the vector-like model, since it depends only on the coupling of the $Z'$ to the SM Higgs.
More constraints on the axial model are discussed in Section~\ref{sec:axial}, whereas the rest of this Section is
devoted to the minimal, vector-like case.

We show below the standard formulae for DM annihilation,
which we reproduce here so as to illuminate the plots we show below~\footnote{In
models where the dark sector contains more than one particle, such as the axial $Y'$-sequential model discussed in Section~\ref{sec:axial}
and the leptophobic model discussed in Section~\ref{sec:leptophobic}, one or more of other `dark' particles
may coannihilate with the dark matter particle. However, this complication
is absent in the vector-like $Y'$-sequential model discussed in this Section, and we neglect it for the other models.}.
Away from the direct-channel $Z^\prime$ and $Z$ resonances, a generic $\chi \chi \to {\bar f} f$ annihilation cross-section multiplied by
the $\chi$ velocity, $\sigma v$, may be expanded as a power series in $v^2$: $\sigma v = a + b v^2 + {\cal O}(v^4)$, where
$a$ and $b$ arise from $s$- and $p$-wave annihilations respectively, and have the following leading-order expressions \cite{Chala:2015ama}
\begin{eqnarray}
a & = & \frac{3 m_\chi^2 \sqrt{1 - \frac{m_f^2}{m_\chi^2}}}{2 \pi (M_{Z^\prime}^2 - 4 m_\chi^2)^2} \left( {g_\chi^V}^2 [ {g_f^V}^2 (2 + \frac{m_f^2}{m_\chi^2}) + 2 {g_f^A}^2 (1 - \frac{m_f^2}{m_\chi^2}) ]+ {g_\chi^A}^2 {g_f^A}^2 \frac{m_f^2}{m_\chi^2} \frac{(4 m_\chi^2 - M_{Z^\prime}^2)^2}{M_{Z^\prime}^4} \right) \nonumber \\ 
b & = & \frac{{g_\chi^A}^2 {g_f^A}^2 m_\chi^2} {2 \pi (M_{Z^\prime}^2 - 4 m_\chi^2)^2} (1 - \frac{m_f^2}{m_\chi^2})^{3/2} \, , \label{ab}
\end{eqnarray}
where $g^{V,A}_{\chi, f}$ are the vector and axial couplings of the dark matter particle and the final-state fermion, respectively. Close to
resonance where $m_\chi \sim M_{Z^\prime}/2$, the denominators in Eq.~(\ref{ab}) are modified: 
$(M_{Z^\prime}^2 - 4 m_\chi^2)^2 \to (M_{Z^\prime}^2 - 4 m_\chi^2)^2 + \Gamma_{Z^\prime}^2 M_{Z^\prime}^2$. As already mentioned,
we include $Z^\prime - Z$ mixing,
and there are analogous modifications when $m_\chi \sim M_{Z}/2$. However care must be taken with the expansion of $\sigma v$ close to resonance, so we always calculate the relic density numerically with {\tt Micromegas}~\cite{micromegas}, with model files generated with {\tt FeynRules}~\cite{feynrules}.

In general, there are regions of any model's parameter space where the relic density exceeds the cold dark matter (CDM)
density inferred from measurements by the Planck satellite and other experiments, $\Omega_{CDM} h^2 \simeq 0.12$. 
We regard these regions as excluded, while noting that modified evolution in the early Universe could change the
calculation of $\Omega_\chi$ so that it is $\le \Omega_{CDM}$, in which case such models could be acceptable~\cite{AEMR}. 
The other generic possibility is that $\Omega_\chi < \Omega_{CDM}$, which
is acceptable if there is some other source of dark matter (for example axions, primordial black holes or sterile neutrinos). However, in this case
the strength of the constraint from the direct search for dark matter scattering is reduced by the density fraction
$\Omega_\chi / \Omega_{CDM}$, a correction that we apply throughout this paper. 
Between these over- and under-denseregions there is a narrow boundary subspace where $\Omega_\chi \simeq \Omega_{CDM}$, and no correction factor is needed.
If $\Omega_\chi > \Omega_{CDM}$, we consider the parameter point to be excluded by the relic density, but for the sake of presenting the direct detection bound we apply no rescaling. 

\begin{figure}
\centering
\includegraphics[width = 0.7\textwidth]{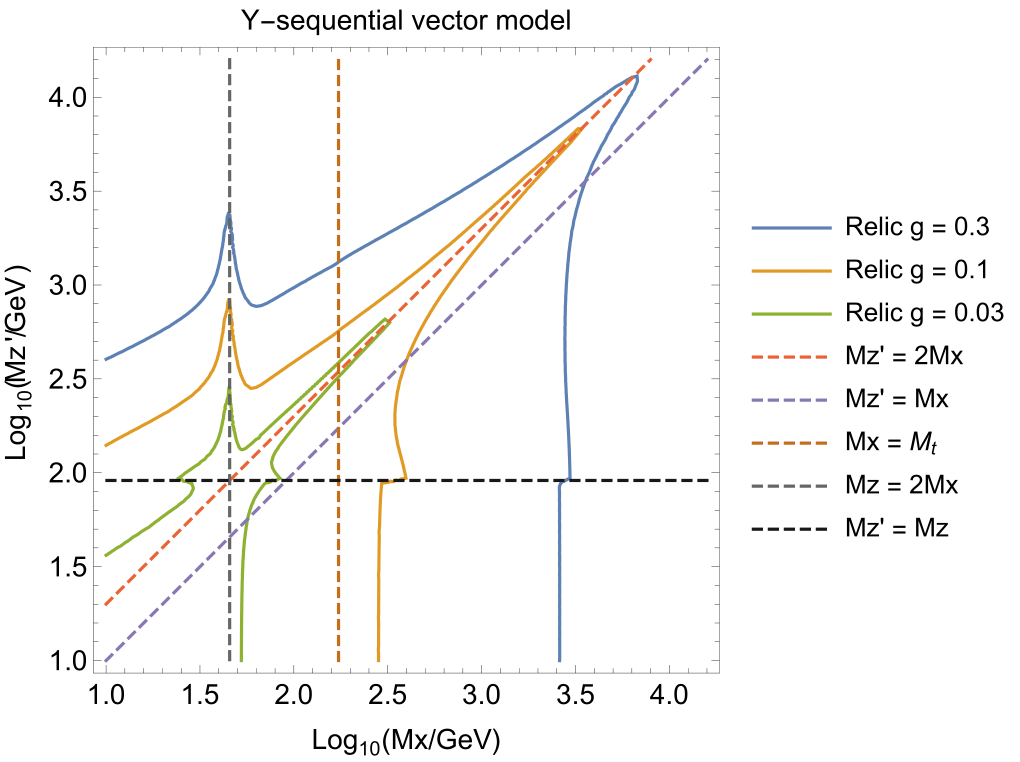}
\caption{\it The $(m_\chi, M_{Z^\prime})$ plane in the U(1)$^\prime$ Y-sequential model with 
a vector-like dark matter coupling $Y'_{\chi,L} = Y'_{\chi,R} = 1$. The
solid lines are contours where $\Omega_\chi = \Omega_{CDM}$ 
for fixed values of the gauge coupling $g = 0.03$ (green), $0.1$ (orange) and $0.3$ (blue), and
the red/purple/brown/grey/black dashed lines are where $m_\chi = M_{Z^\prime}/2, m_\chi = M_{Z^\prime}$, $m_\chi = m_t$,
$m_\chi = M_Z/2$, $M_{Z'} = M_Z$, respectively.}
\label{fig:VectorOmega}
\end{figure}

Fig.~\ref{fig:VectorOmega} displays this boundary in the $(m_\chi, M_{Z^\prime})$ plane in the vector-like $Y^\prime$-sequential model.
The solid contours are at the boundaries where $\Omega_\chi = \Omega_{CDM}$,
for $Y'_{\chi,L} = Y'_{\chi,R} = 1$ and fixed values of the gauge coupling $g = 0.03$ (green curve), 0.1 (orange) and 0.3 (blue)~\footnote{The 
curves for other parameter choices with the same values of
$g^2 |Y'_{\chi,L}|$ would be identical away from resonance.}.
The narrow-width approximation assumed in our
analysis would no longer be applicable for $g > 0.3$, so we do not display results for larger $g$.
Also shown are dashed lines where $m_\chi = M_{Z^\prime}/2$ (red), $m_\chi = M_{Z^\prime}$ (purple)
and $m_\chi = m_t$ (brown), $m_\chi = M_Z/2$ (grey) and $M_{Z'} = M_Z$ (black). 

The relic $\chi$
density is reduced below the relic CDM density by rapid annihilation $\chi \chi \to Z^\prime \to \text{SM SM}$, for SM a Standard model particle,
in wedge-shaped regions where $m_\chi \sim M_{Z^\prime}/2$, whose widths increase with $g$. 
Larger values of $\Omega_\chi$ arise when 
the dark matter annihilation rate decreases as $M_{Z^\prime}$ increases.
We note that the wedge-shaped contours exhibit outward-pointing glitches when
$m_\chi \simeq M_Z/2$, where the relic density is suppressed by rapid $\chi \chi$ annihilations
via the $Z$ to Standard Model particles, and when $M_{Z'} \simeq M_Z$, where $Z^\prime - Z$ mixing is enhanced. 
As already mentioned, for
parameter sets inside the wedges we rescale the constraint from the direct dark matter scattering rate by a factor
$\Omega_\chi / \Omega_{CDM}$, whereas the regions outside these wedges are
disallowed because $\Omega_\chi > \Omega_{CDM}$. 

\begin{figure}
\centering
\includegraphics[width = 0.45\textwidth]{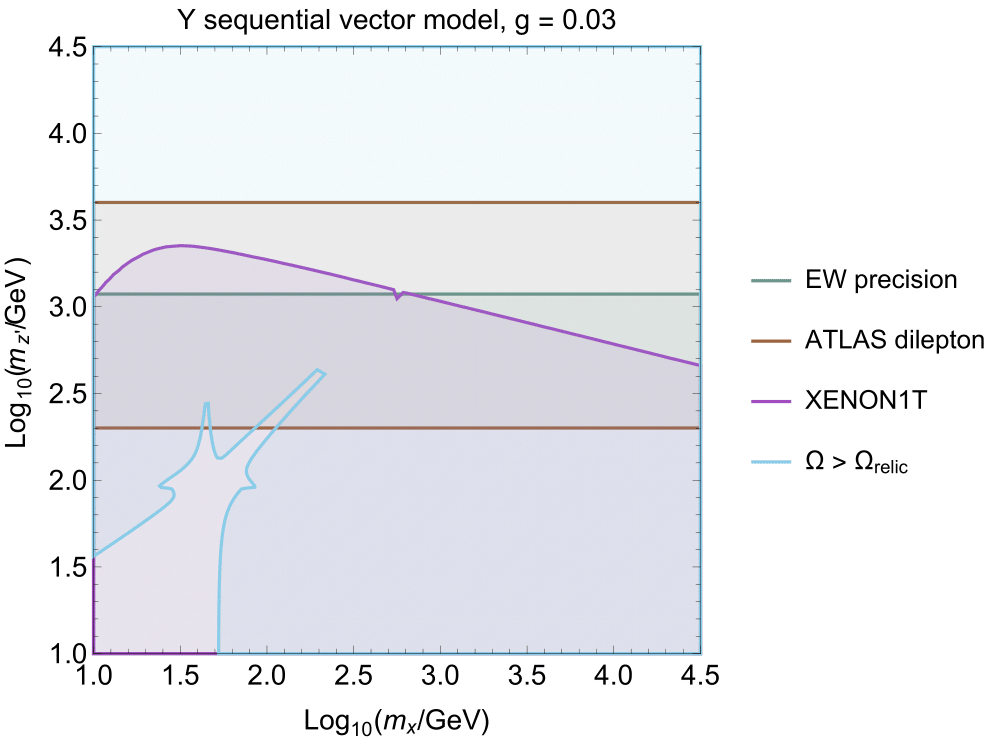}
\hspace{0.5cm}
\includegraphics[width = 0.45\textwidth]{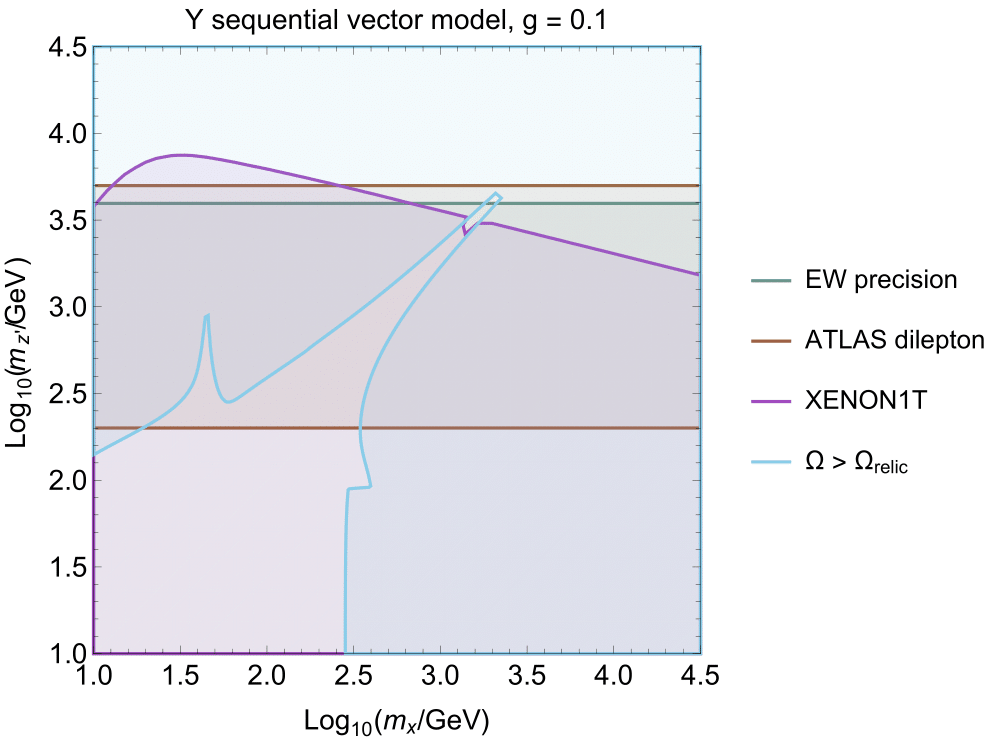}\\
\vspace{0.5cm}
\includegraphics[width = 0.45\textwidth]{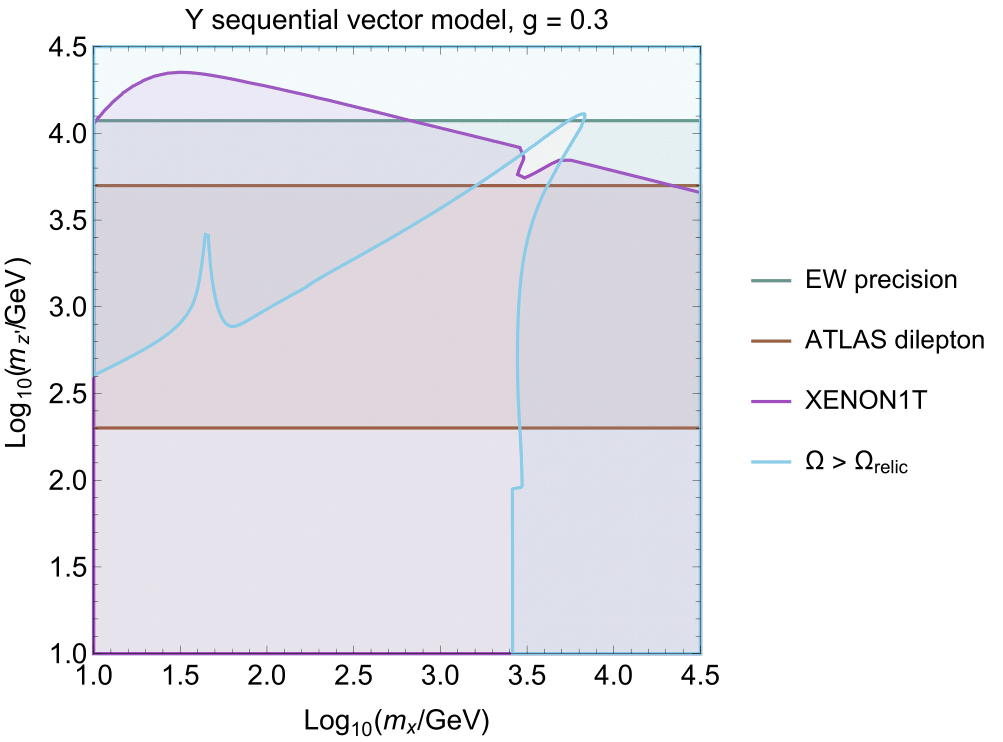}
\vspace{0.5cm}
\caption{\it The $(m_\chi, M_{Z^\prime})$ planes in the U(1)$^\prime$ Y-sequential model with 
a vector-like dark matter coupling $Y'_{\chi,L} = Y'_{\chi,R} = 1$ for a gauge coupling $g = 0.03$ (upper left),
$g = 0.1$ (upper right) and $g = 0.3$ (lower). The
solid blue lines are the contours where $\Omega_\chi = \Omega_{CDM}$, and $\Omega_\chi > \Omega_{CDM}$
in the regions shaded blue. The bands shaded orange are excluded by the ATLAS dilepton search~\cite{LHCZprime}, the regions shaded olive
are excluded by precision electroweak measurements, and the direct XENON1T constraint~\cite{XENON1T} on dark matter scattering
are shown as purple lines.}
\label{fig:VectorFixedg}
\end{figure}

Fig.~\ref{fig:VectorFixedg} displays $(m_\chi, M_{Z^\prime})$ planes in the vector-like U(1)$^\prime$ Y-sequential model
for three selected values of the gauge coupling: $g = 0.03$ (upper left), $g = 0.1$ (upper right) and $g = 0.3$ (lower),
implementing the following constraints.
In each panel the blue contour is where $\Omega_\chi = \Omega_{CDM}$ and the blue shaded regions are excluded
because $\Omega_\chi > \Omega_{CDM}$. The horizontal olive lines at fixed values of $M_{Z'}$ that rise with
increasing $g$ bound the olive shaded regions at lower $M_{Z'}$ that are
excluded by the constraints imposed by precision electroweak measurements induced by the effects of $Z' - Z$ mixing . 
For large $g \gtrsim 0.1$, this constraint is stronger than the ATLAS
dilepton search at the LHC~\cite{LHCZprime}, which excludes the orange shaded regions~\footnote{We have used {\tt MadGraph}~\cite{madgraph} to calculate the dilepton, dijet, and monojet constraints.}. Finally, the purple shaded regions are 
excluded by the direct search for the scattering of dark matter by the XENON1T experiment~\cite{XENON1T}, 
where the appropriate reduction factor $\Omega_\chi / \Omega_{CDM}$ has been applied to the
experimental upper limit.  Here and throughout we use the approximations in~\cite{ddtools} to calculate direct detection limits.  In the $g = 0.03$ and $0.1$ cases there is no visible region that is allowed by all these constraints.
On the other hand, when $g = 0.3$ we see a tiny region that is only just consistent with the relic density 
and precision electroweak constraints, while being more comfortably consistent with
the dark matter scattering and ATLAS dilepton constraints.

\begin{figure}
\includegraphics[width = 0.475\textwidth]{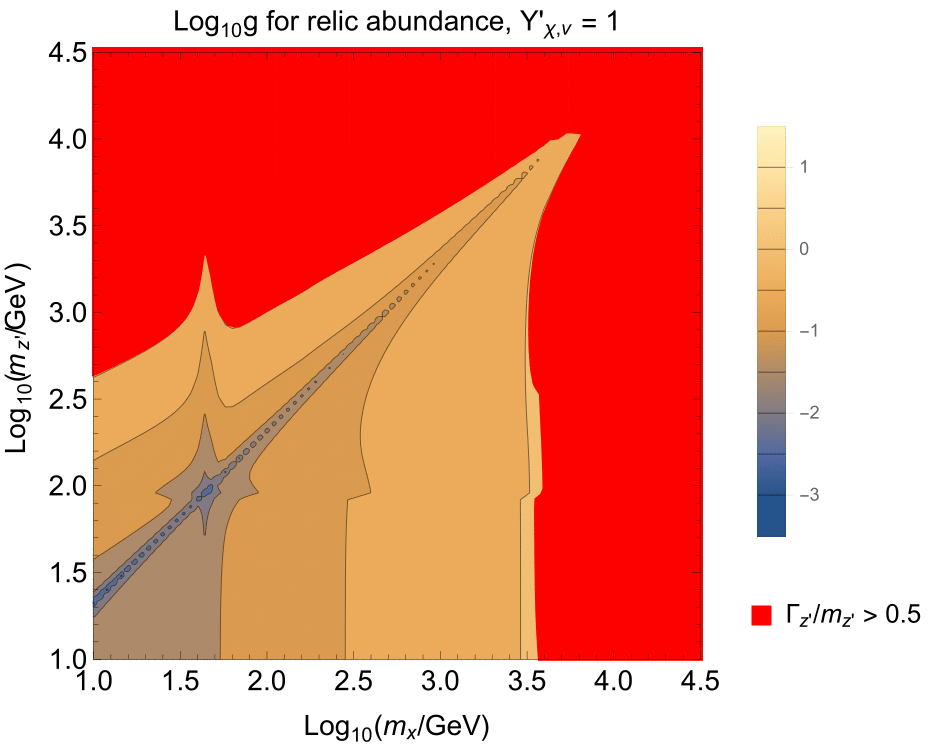}
\includegraphics[width = 0.525\textwidth]{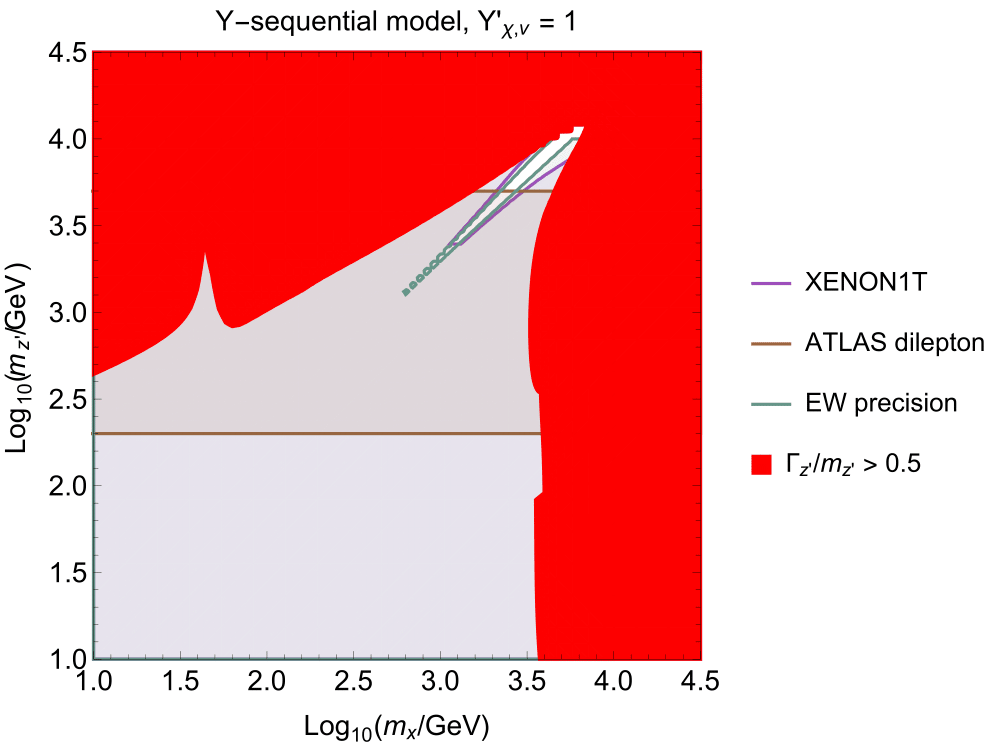}
\caption{\it Left panel: The $(m_\chi, M_{Z^\prime})$ plane in the U(1)$^\prime$ Y-sequential model with 
a vector-like dark matter coupling $Y'_{\chi,L} = Y'_{\chi,R} = 1$ and the value of the gauge coupling $g$ allowed to vary
so as to yield $\Omega_\chi = \Omega_{CDM}$ everywhere in the plane. 
Right panel: The same $(m_\chi, m_{Z'})$ plane with varying gauge coupling $g$, now showing
the band excluded by the ATLAS dilepton search (shaded orange)~\cite{LHCZprime}, the regions excluded by the direct XENON1T
search for dark matter scattering (shaded purple)~\cite{XENON1T}, the region excluded by precision electroweak data
(shaded olive)~\cite{sandt2018} and the regions where  $\Gamma_{Z^\prime}/M_{Z^\prime} > 0.5$ (shaded red).
Note the small allowed region with $\log_{10} (m_\chi/{\rm GeV}) \gtrsim 3.3$
and $\log_{10} (M_{Z^\prime}{\rm GeV}) \gtrsim 3.7$.  Note that
there is a very narrow region on resonance that is not visible on this plot but is explored in Fig.~\ref{fig:Diagonal}}
\label{fig:VectorVaryg}
\end{figure}

Finally, we present in Fig.~\ref{fig:VectorVaryg} an analysis of the vector-like 
U(1)$^\prime$ Y-sequential model in which $g$ is varied 
so as to maintain $\Omega_\chi = \Omega_{CDM}$ across the $(m_\chi, M_{Z^\prime})$ plane. 
In the left panel the values of $g$ required by the relic density
are indicated by the indicated shadings, and the red shaded regions correspond to
$\Gamma_{Z'}/M_{Z'} > 0.5$ which are excluded from our analysis.
The right panel of Fig.~\ref{fig:VectorVaryg} shows the interplay of
the LHC (brown shading) and dark matter search (purple) constraints in this plane with varying $g$, 
as well as the precision electroweak constraint (olive). As in the left panel, we exclude the red shaded regions where
$\Gamma_{Z^\prime}/M_{Z^\prime} > 0.5$. There is a visible area where the vector-like U(1)$^\prime$ Y-sequential model is
compatible with all the constraints, in the high mass resonance region where $M_{Z^\prime} \approx 2 m_\chi$ around $\log_{10} (m_\chi / {\rm GeV}) \gtrsim 3.3$
and $\log_{10} (M_{Z^\prime} / {\rm GeV}) \gtrsim 3.7$. This small allowed region at larger $\log_{10} (m_\chi/{\rm GeV}) \sim 3.5$
and $\log_{10} (M_{Z^\prime}/{\rm GeV}) \gtrsim 3.7$ is squeezed by the requirement that $\Gamma_{Z^\prime}/M_{Z^\prime} < 0.5$.

\begin{figure}
\includegraphics[width = 0.5\textwidth]{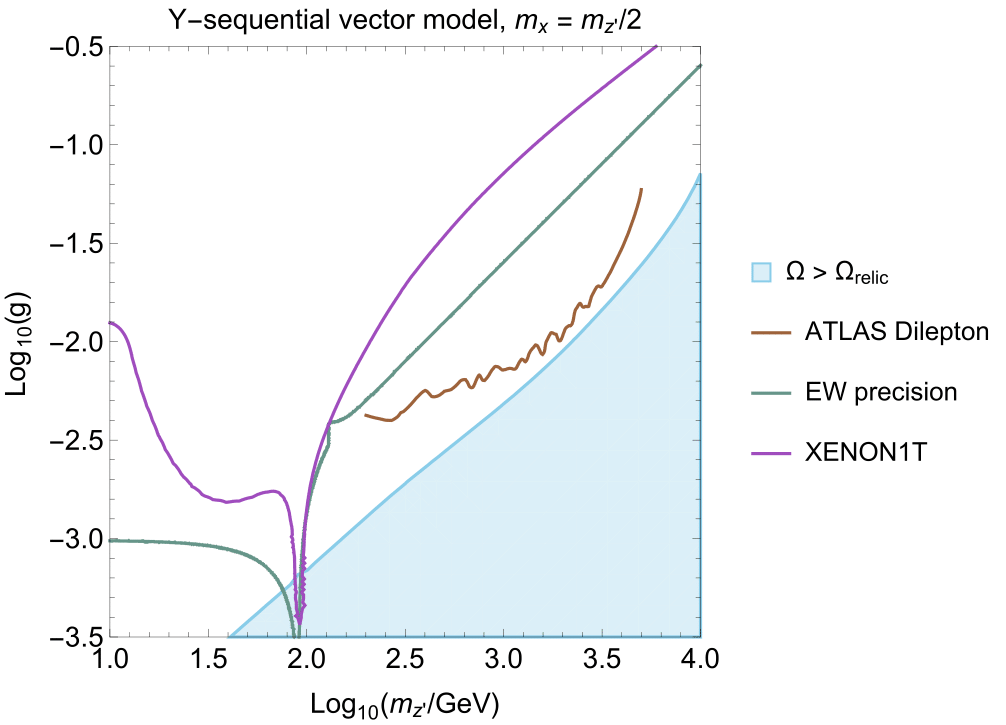}
\includegraphics[width = 0.5\textwidth]{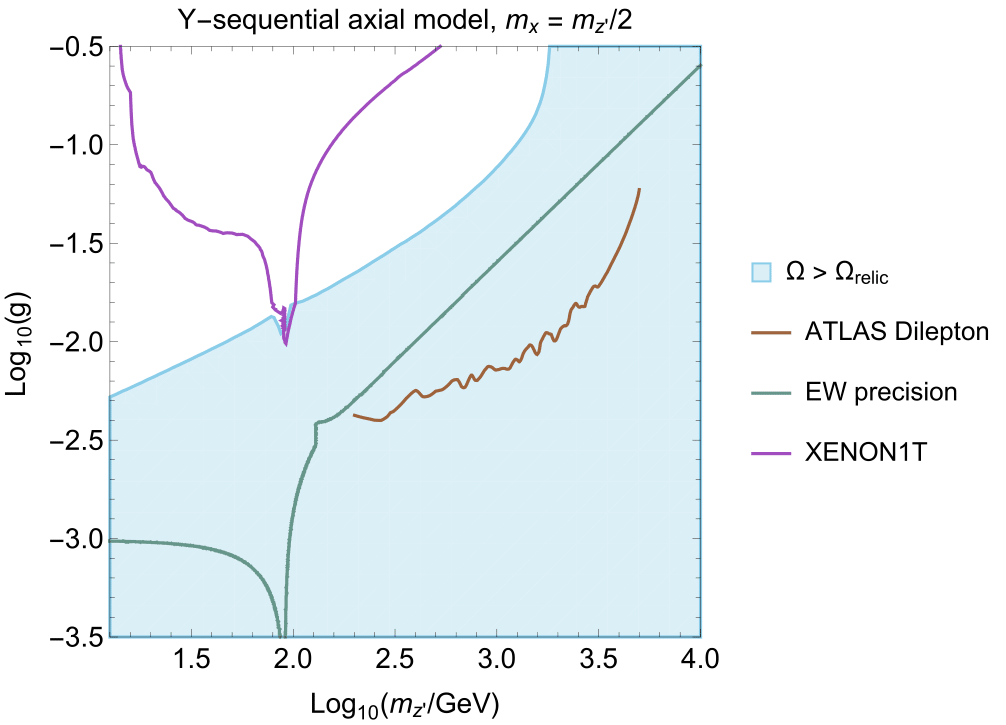}
\caption{\it The interplays of the constraints on the vector-like (left panel) 
and axial (right panel) U(1)$^\prime$ Y-sequential models
along the line $m_\chi = M_{Z^\prime}/2$, for a range of values of the gauge coupling $g$.
The relic density $\Omega_\chi = \Omega_{CDM}$ along the blue lines, and the relic density is too high
in the blue-shaded region below it. The ATLAS dilepton constraint~\cite{LHCZprime} is shown as brown lines: regions above
are excluded. The purple lines are the upper limits on $g$ from direct dark matter searches, and
the green lines show the upper bound from precision electroweak data.}
\label{fig:Diagonal}
\end{figure}

In addition to this visible allowed region, there is also a narrow sliver of parameter space where $M_{Z'} \sim 2 m_\chi$
that is also compatible with all the constraints, which is invisibly thin in Fig.~\ref{fig:VectorVaryg} \cite{fairbairnheal}.
The left panel of Fig.~\ref{fig:Diagonal} displays the relevant constraints on the vector-like U(1)$^\prime$ Y-sequential model
along the line $M_{Z'} = 2 m_\chi$ for a range of values of $g$. The relic density 
$\Omega_\chi = \Omega_{CDM}$ along the blue line, and the blue-shaded region below it is excluded because 
the relic particle is overabundant. The ATLAS dilepton constraint~\cite{LHCZprime} is shown as a brown line extending over the range
$2.3 \lesssim \log_{10} (M_{Z^\prime} / {\rm GeV}) \lesssim 3.7$, with the region above being excluded.
The purple line shows the upper limit on $g$ provided by direct dark matter searches as a function of $M_{Z^\prime}$.
Finally, the green line reproduces the constraint  from precision electroweak data. We see that there is a region to the right of this
line, below the direct search and ATLAS dilepton lines and above the blue line that is compatible with all the
constraints. Points above the blue line would have $\Omega_\chi < \Omega_{CDM}$, but the relic density could be
brought up to the limit $\Omega_\chi = \Omega_{CDM}$ by taking $m_\chi$ slightly below or above $ M_{Z'}/2 $, so that the 
$\chi \chi$ annihilation cross-section is suitably reduced by sliding down one of the sides of the $Z'$ Breit-Wigner peak.

The conclusion of this analysis of the vector-like U(1)$^\prime$ Y-sequential model is similar to what was foreseen in~\cite{usEFT1}.
It is very tightly constrained by the ATLAS dilepton search and direct searches for dark matter scattering, as well as the precision electroweak data, with the only allowed region (apart from the very narrow resonance region
discussed in the previous paragraph) appearing when with $\log_{10} (m_\chi/{\rm GeV}) \gtrsim 3.3$
and $\log_{10} (M_{Z^\prime}{\rm GeV}) \gtrsim 3.7$. 

\section{Benchmark with an Axial Dark Matter Coupling\label{sec:axial}}

In this Section we consider another variant of the $Y'$-sequential model, assuming again that any exotic fermions are SM singlets such that the U(1)$^\prime$ charges of the Standard Model particles are guaranteed to be:
\begin{equation}
Y'_q \; = \; 1, \quad Y'_l \; = \; -3, \quad Y'_e \; = \; -6, \quad Y'_d \; = \; -2 , \quad Y'_u \; = \; 4 ,  \quad Y'_H \; = \; -3\, .
\label{Bench2}
\end{equation}
However, in contrast to the previous Section, we now consider a case
where the dark matter particle $\chi$ has an axial U(1)$^\prime$ coupling: $Y'_{\chi,L} = - Y'_{\chi,R}$.
As in the vector-like case discussed in the previous Section, the model has as free parameters the
U(1)$^\prime$ coupling $g$, $m_\chi$, $M_{Z^\prime}$ and the magnitude of the U(1)$^\prime$ charge of the 
dark matter particle. In addition, this benchmark must have at least one additional dark sector particle so as
to cancel the triangle anomalies, as discussed in~\cite{usEFT1}. However, here we do not discuss further the
possible phenomenology of such an extended dark sector.

\begin{figure}
\centering
\includegraphics[width = 0.5\textwidth]{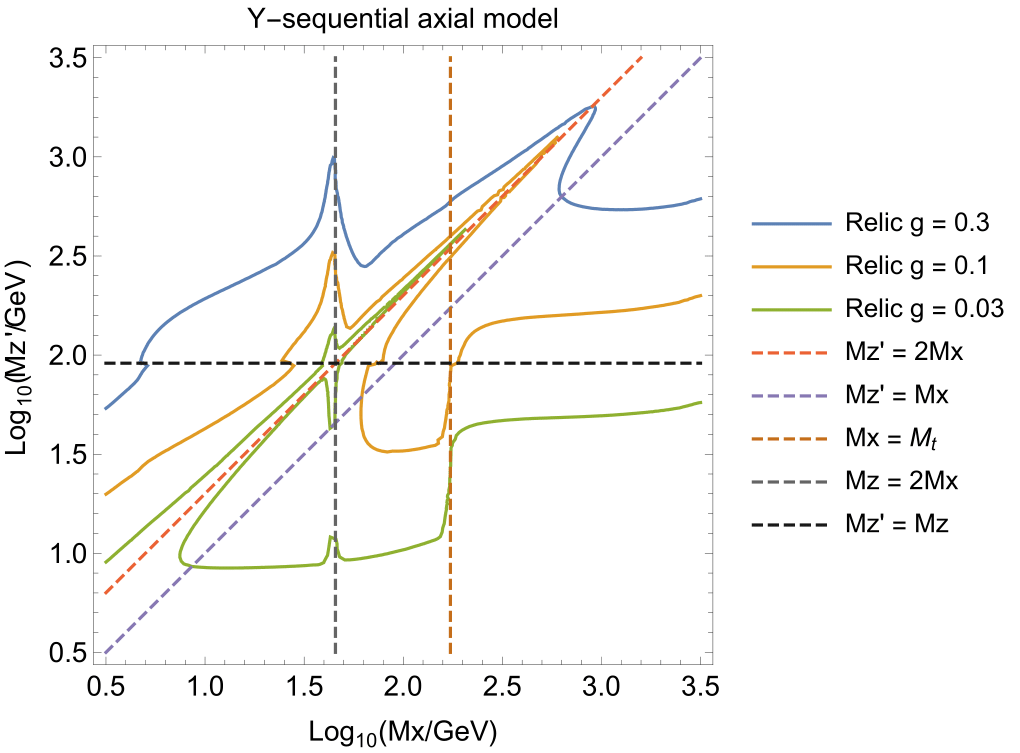}
\caption{\it The $(m_\chi, M_{Z^\prime})$ plane in the U(1)$^\prime$ Y-sequential model with 
an axial-like dark matter coupling $Y'_{\chi,L} = - Y'_{\chi,R} = 1$. The
solid green (orange) (blue) lines are the contours where $\Omega_\chi = \Omega_{CDM}$ for $g = 0.03 (0.1) (0.3)$, and
the red/purple/brown/grey/black dashed lines are where $m_\chi = M_{Z^\prime}/2, m_\chi = M_{Z^\prime}$, $m_\chi = m_t$,
$m_\chi = M_{Z}/2$ and $M_{Z'} = M_Z$, respectively.}
\label{fig:AxialOmega}
\end{figure}

Fig.~\ref{fig:AxialOmega} displays the $(m_\chi, M_{Z^\prime})$ plane in this model with 
$Y'_{\chi,L} = - Y'_{\chi,R} = 1$, analogous to the vector-like case shown in Fig.~\ref{fig:VectorOmega}.
We show as solid green (orange) (blue) lines the contours where $\Omega_\chi = \Omega_{CDM}$
for the same choices $g = 0.03 (0.1) (0.3)$ considered above~\footnote{We
recall that, away from resonance,
these contours would be similar for other axial models with the same value of $g^2 |Y'_{\chi,L}|$.}, and the relations
$m_\chi = M_{Z^\prime}/2, m_\chi = M_{Z^\prime}$, $m_\chi = m_t$, $m_\chi = M_{Z'}/2$
and $M_{Z'} = M_Z$ are again shown by dashed lines with the same colours as in the vector-like case.
As in that case, the dark matter contour exhibits a wedge around $m_\chi = M_{Z^\prime}/2$, which is asymmetric and
extends to large $m_\chi$ when $\log_{10} (M_{Z^\prime} / {\rm GeV}) \lesssim 2$. 
This extension is due to the opening of the
$\chi \chi \to {\bar t} t$ threshold when $m_\chi > m_t$. Below this threshold, annihilations into pairs of Standard Model fermions
are suppressed by mass factors (helicity suppressed), as can be seen in the second line of Eq.~(\ref{ab}).
For this reason, the dominant $\chi \chi$ annihilation channel is into pairs of mediator bosons, $Z^\prime Z^\prime$,
when $m_t \gtrsim m_\chi \gtrsim M_{Z^\prime}$. The relic density contours also exhibit glitches associated with
enhanced annihilation when $\chi \chi \to Z$ on resonance, induced by $Z - Z'$ mixing,
and when $M_{Z'} \simeq M_Z$ this mixing is enhanced.

\begin{figure}
\centering
\includegraphics[width = 0.45\textwidth]{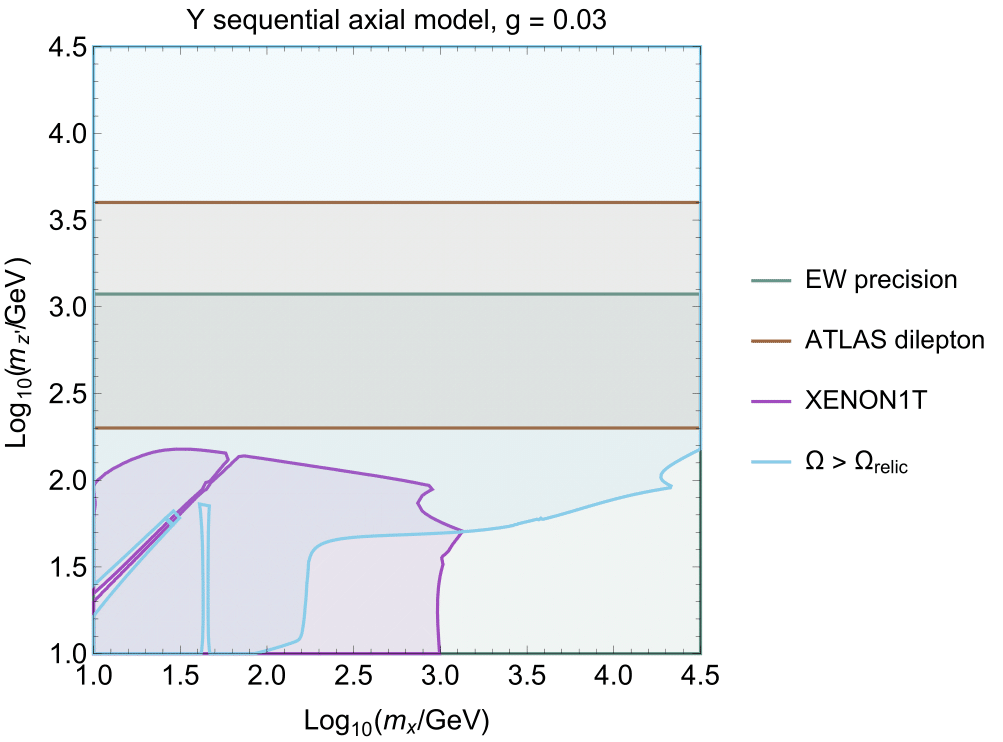}
\includegraphics[width = 0.45\textwidth]{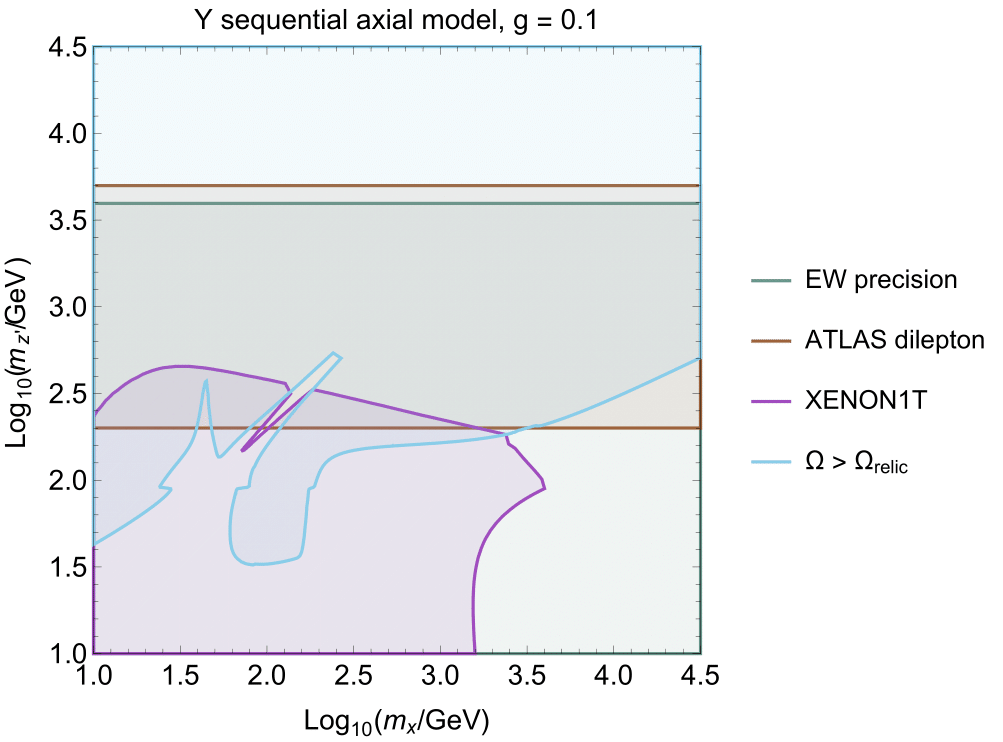}\\
\includegraphics[width = 0.45\textwidth]{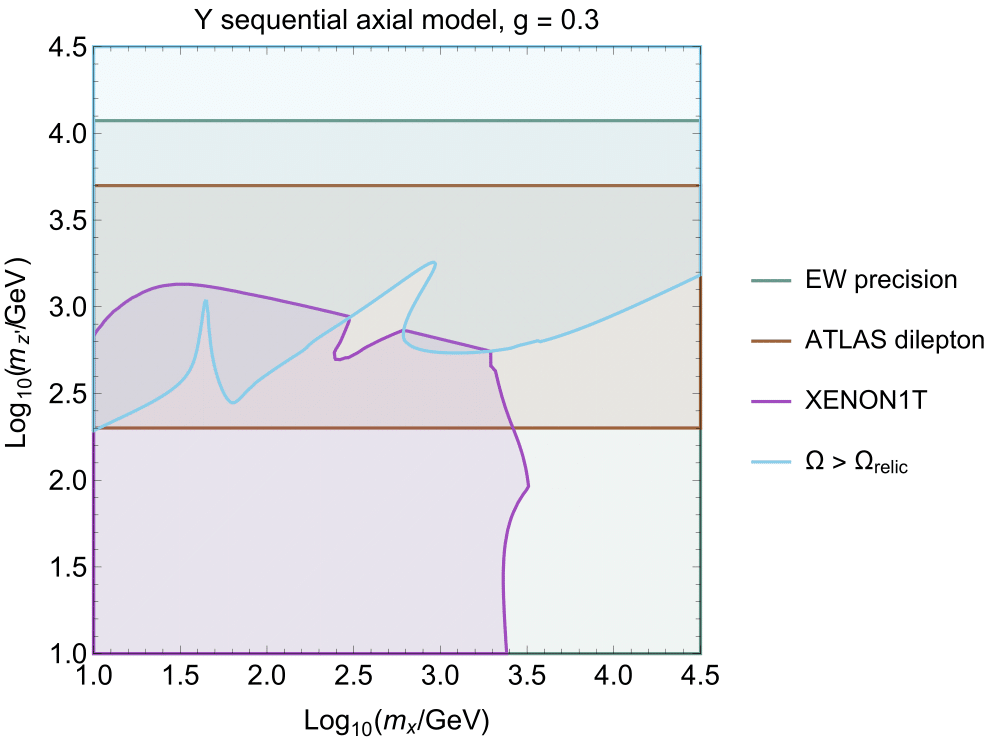}
\caption{\it The $(m_\chi, M_{Z^\prime})$ planes in the U(1)$^\prime$ Y-sequential model with 
an axial-like dark matter coupling $Y'_{\chi,L} = - Y'_{\chi,R} = 1$ for a gauge coupling $g = 0.03$ (upper left),
$g = 0.1$ (upper right) and $g = 0.3$ (lower). The
solid blue lines are the contours where $\Omega_\chi = \Omega_{CDM}$, and $\Omega_\chi > \Omega_{CDM}$
in the regions shaded blue. The bands shaded brown are excluded by the ATLAS dilepton search, the regions shaded purple
are excluded by direct searches for dark matter scattering, and the regions shaded olive are excluded by precision electroweak data.}
\label{fig:AxialFixedg}
\end{figure}

Fig.~\ref{fig:AxialFixedg} displays the $(m_\chi, M_{Z^\prime})$ planes in the axial U(1)$^\prime$ Y-sequential model
for the following fixed values of $g$, assuming $Y'_{\chi,L} = - Y'_{\chi,R} = 1$: $g = 0.03$ (upper left), 0.1 (upper right)
and 0.3 (lower). As in the vector-like case, we do not consider larger values of $g$, because the narrow-width
approximation for the $Z^\prime$ breaks down. As in Fig.~\ref{fig:VectorFixedg}, the regions of the planes where
$\Omega_\chi > \Omega_{CDM}$ are shaded blue, those excluded by the ATLAS dilepton search are shaded
brown, those excluded by the (suitably rescaled) direct dark matter searches are shaded purple, and those excluded
by precision electroweak measurements are shaded green.
When $g = 0.03$ and $0.1$ the dark matter density constraint is in general more powerful than the ATLAS constraint,
but they are more complementary when $g = 0.3$.
The direct dark matter search constraint is important at low $m_\chi$ and $M_{Z'}$,
and is relatively similar in the three panels, strengthening slightly
as $g$ increases. However, the most important constraint for $M_{Z'} \lesssim 3$ to $4$~TeV
is that from precision electroweak data. In combination with the relic density constraint that excludes larger $M_{Z'}$,
it excludes all the displayed region of the $(m_\chi, M_{Z^\prime})$ plane.

\begin{figure}
\includegraphics[width = 0.475\textwidth]{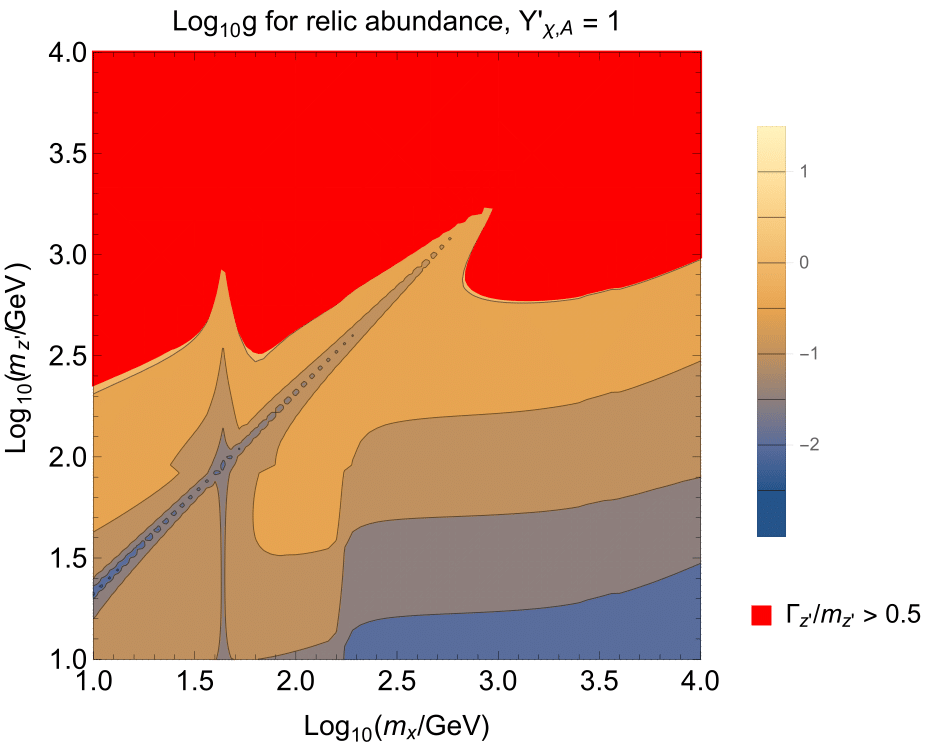}
\includegraphics[width = 0.525\textwidth]{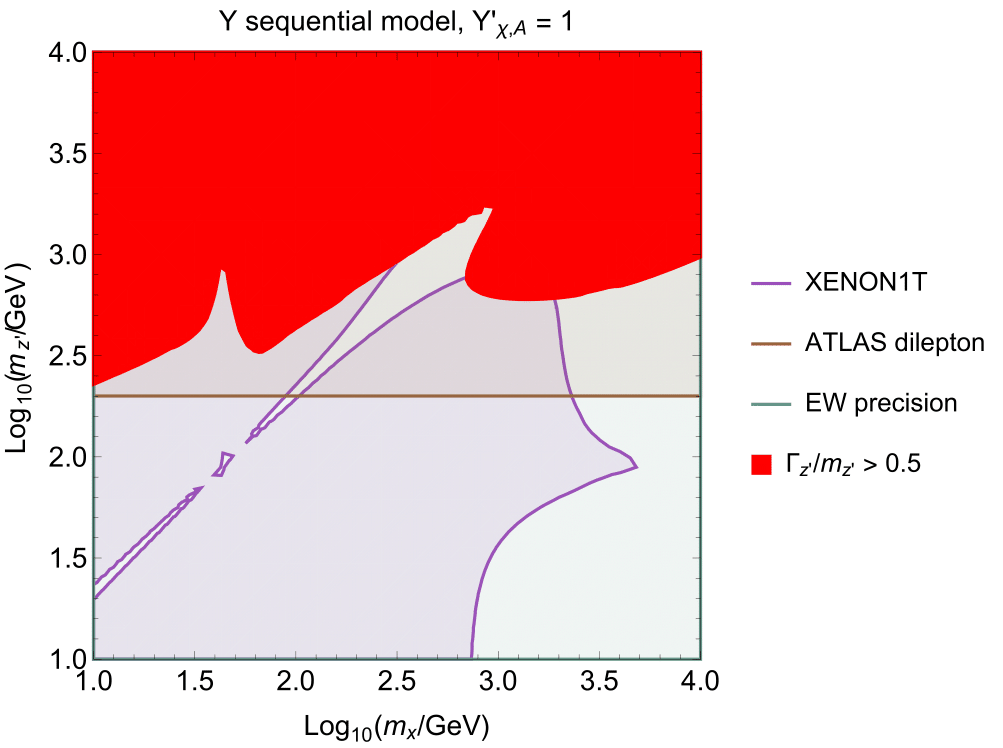}
\caption{\it Left panel: The $(m_\chi, M_{Z^\prime})$ plane in the U(1)$^\prime$ Y-sequential model with 
an axial-like dark matter coupling $Y'_{\chi,L} = - Y'_{\chi,R} = 1$ and the value of the gauge coupling $g$ chosen
to yield $\Omega_\chi = \Omega_{CDM}$. Right panel: The $(m_\chi, m_{Z'})$ plane with this varying gauge coupling $g$, showing
the band excluded by the ATLAS dilepton search (shaded orange), the regions excluded by direct searches for dark matter scattering
(shaded purple), the precision electroweak constraints (olive) and the regions where  $\Gamma_{Z^\prime}/M_{Z^\prime} > 0.5$ (red).
None of the displayed region of the plane is consistent with all the constraints since the electroweak constraints rule out the entire plane.}
\label{fig:AxialVaryg}
\end{figure}

We show in the left panel of Fig.~\ref{fig:AxialVaryg} the $(m_\chi, M_{Z^\prime})$ plane in the axial U(1)$^\prime$ Y-sequential model
with $g$ allowed to vary as indicated by the colour coding shown in the legend, 
so as to obtain $\Omega_\chi = \Omega_{CDM}$ throughout the plane. 
As in the vector-like case shown in the left panel of Fig.~\ref{fig:VectorVaryg}, there is a region at large $M_{Z^\prime}$
where the required value of $g$ becomes large and even non-perturbative. Shaded in red is the region where
$\Gamma_{Z'}/M_{Z'} > 0.5$. One difference from the vector-like case is the series of
`steps' in the contours of $g$ at $\log_{10} (m_\chi / {\rm GeV}) \sim 2.7$ where the onset of the ${\bar t} t$ threshold increases the
annihilation rate for fixed $g$, so that a smaller value of $g$ is needed to obtain $\Omega_\chi = \Omega_{CDM}$. 

This feature is
reflected in the right panel of Fig.~\ref{fig:AxialVaryg}, where we see that the exclusion by the direct search for dark matter scattering
(purple shading) runs out of steam when $\log_{10} (m_\chi / {\rm GeV}) \gtrsim 3$ and $g$ is small. For
the same reason, it is also weakened along the diagonal
line where $m_\chi \simeq M_{Z'}/2$. We also note the region at large $M_{Z^\prime}$
where $\Gamma_{Z^\prime} / M_{Z^\prime} > 0.5$, and that the ATLAS dilepton constraint
again enforces $\log_{10} (M_{Z^\prime} / {\rm GeV}) \lesssim 2.3$.
Lower values of $M_{Z^\prime}$ are excluded by the precision electroweak data, as in the vector-like case.

However, we see in the right panel of Fig.~\ref{fig:AxialVaryg} that
there is no part of the displayed region of the $(m_\chi, M_{Z^\prime})$ plane
in the Y-sequential model with an axial $Z'$ dark matter coupling that is consistent
with all the constraints. In particular, in this instance, unlike in the vector-like case, there is no allowed strip when 
$m_\chi \simeq M_{Z'}/2$, as there was in the left panel of Fig.~\ref{fig:Diagonal}. This is mainly a result of the fact that the 
annihilation cross section is p-wave suppressed (resulting in a $v^2$ suppression), 
which requires the gauge coupling $g$ to be larger to match the observed value of the relic density.

\section{Benchmarks with a Leptophobic $Z^\prime$\label{sec:leptophobic}}

We now consider two benchmark leptophobic models
that were also originally proposed in~\cite{usEFT1}. By construction, 
they both have $Y'_l = Y'_e = 0$, which is possible only if there are additional particles beyond the dark matter particle.
The first model we study contains an additional SU(2) doublet of fermions $B$.
In the visible sector it has universal U(1)$^\prime$ charges for the quarks:
\begin{equation}
Y'_q \; = \; Y'_u \; = Y'_d \, ,
\label{Leptophobic1}
\end{equation}
and hence $Y'_H = 0$. Normalizing the U(1)$^\prime$ coupling so that $Y'_{\chi, L} = 1$,
the following are the U(1)$^\prime$ charges of the quarks and the $\chi_R$:
\begin{equation}
Y'_q \; = \; -\frac{1}{27}, \quad Y'_{\chi,R} \; = \; 0 \, ,
\label{Charges1}
\end{equation}
and the U(1)$^\prime$ charges of the left- and right-handed components of the additional SU(2) doublet $B$ are
\begin{equation}
Y'_{B,L} \; = \; -\frac{1}{3}, \quad Y'_{B,R} \; = \; \frac{4}{3} \, .
\label{Bcharges1}
\end{equation}
The leptophobia of this model implies that the ATLAS dilepton search constraint is irrelevant. However, one must still consider
the (weaker) constraint from searches for structures in the dijet spectrum. In addition, 
the small size of the quark charges in Eq.~(\ref{Charges1})
compared to the charge of the dark matter particle implies that the LHC monojet + $\ETslash$ constraint is also important.
The absence of leptonic U(1)$^\prime$ charges
implies that the Higgs multiplet must also have vanishing $Y^\prime$, which implies that tree-level $Z - Z^\prime$ mixing through the
Higgs sector is absent. However, the presence of particles with both Standard Model and U(1)$^\prime$ charges implies that
kinetic $Z - Z^\prime$ mixing is induced at the loop level (we assume $\epsilon = 0$ at tree level), as we discuss in Appendix~\ref{AppMix}.

The second leptophobic $Z'$ model that we consider contains instead an additional SU(2) triplet of fermions. 
It has the following universal U(1)$^\prime$ charges for the quarks:
\begin{equation}
Y'_{q} = - \frac{2}{9}, \quad Y'_H \; = \; 0 \, ,
\label{Charges2}
\end{equation}
where we have again normalized the U(1)$^\prime$ coupling so that $Y'_{\chi,L} =1$,
and vanishing Higgs charge. In addition, this model has $Y'_{\chi,R} = 1/2$ and
the following charges for the left- and right-handed components of the additional SU(2) triplet:
\begin{equation}
Y'_{B,L} = - \frac{1}{2}, \quad Y'_{B,R} = \frac{1}{2} \, .
\label{Bcharges2}
\end{equation}
In this model the quark charges in Eq.~(\ref{Charges2}) are less suppressed relative to
the charge of the dark matter particle than in the first leptophobic benchmark model, so that the LHC monojet + $\ETslash$ constraint is correspondingly less important.

\begin{figure}
\includegraphics[width = 0.5\textwidth]{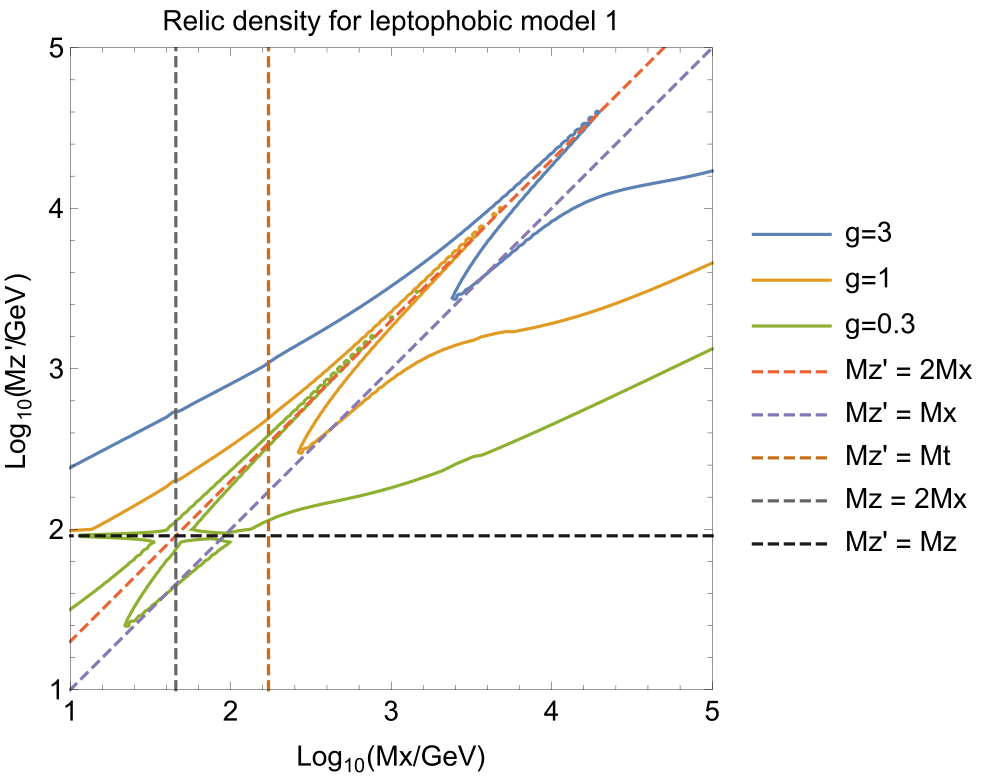}
\includegraphics[width = 0.5\textwidth]{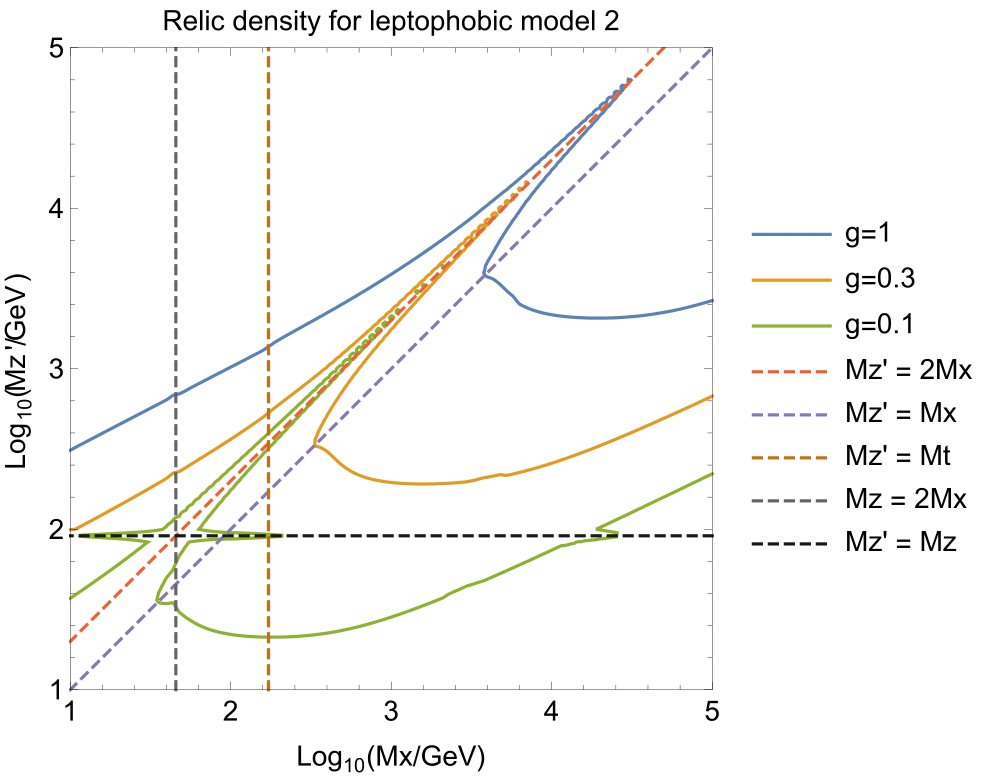}
\caption{\it The $(m_\chi, M_{Z^\prime})$ planes in the leptophobic U(1)$^\prime$
models (Eq.~(\protect\ref{Charges1}) (left panel) and Eq.~(\protect\ref{Charges2}) (right panel)).
The solid lines are contours where $\Omega_\chi = \Omega_{CDM}$ for the indicated
choices of the U(1)$^\prime$ coupling $g$, and the red/purple/orange/grey/black dashed lines are
where $m_\chi = M_{Z'}/2, m_\chi = M_{Z'}, m_\chi =m_t, m_\chi = M_Z$ and $M_{Z'} = M_Z$, respectively.
}
\label{fig:leptophobicfixedg}
\end{figure}
Fig.~\ref{fig:leptophobicfixedg} displays in the left panel the $(m_\chi, M_{Z^\prime})$ plane in the 
first leptophobic U(1)$^\prime$ model with the U(1)$^\prime$ charges shown in Eq.~(\ref{Charges1}),
and in the right panel plane the corresponding $(m_\chi, M_{Z^\prime})$ plane in the 
second leptophobic U(1)$^\prime$ model (Eq.~(\ref{Charges2})). The solid lines are contours 
where $\Omega_\chi = \Omega_{CDM}$ for the indicated fixed choices of the U(1)$^\prime$ coupling $g$.
The choices of $g$ are different because the larger quark U(1)$^\prime$ charges in the second
model imply that its total decay width is larger than in the first model for the same value of $g$,
causing the narrow-width approximation to break down for a smaller value of $g$ than is the case
in the first leptophobic model Eq.~(\ref{Charges1}).

In both cases, we see the familiar feature that larger values of $m_\chi$ and $M_{Z^\prime}$
are compatible with the $\Omega_\chi = \Omega_{CDM}$ constraint along the dashed red diagonal
line where $m_\chi = M_{Z'}/2$. Below this diagonal line, the contours in the two models are quite different when
$m_\chi > M_{Z'}$ (below and to the right of the diagonal purple dashed line), reflecting the
greater importance of $\chi \chi$ annihilations into pairs of $Z'$ bosons relative to annihilations into SM particles. This is because the first leptophobic
model has a smaller quark U(1)$^\prime$ charge (shown in Eq.~(\ref{Charges1}) while the dark matter charges are somewhat similar.
We also note glitches in the relic density contours where $M_{Z'} = M_Z$ (black dashed lines).

\begin{figure}
\includegraphics[width = 0.5\textwidth]{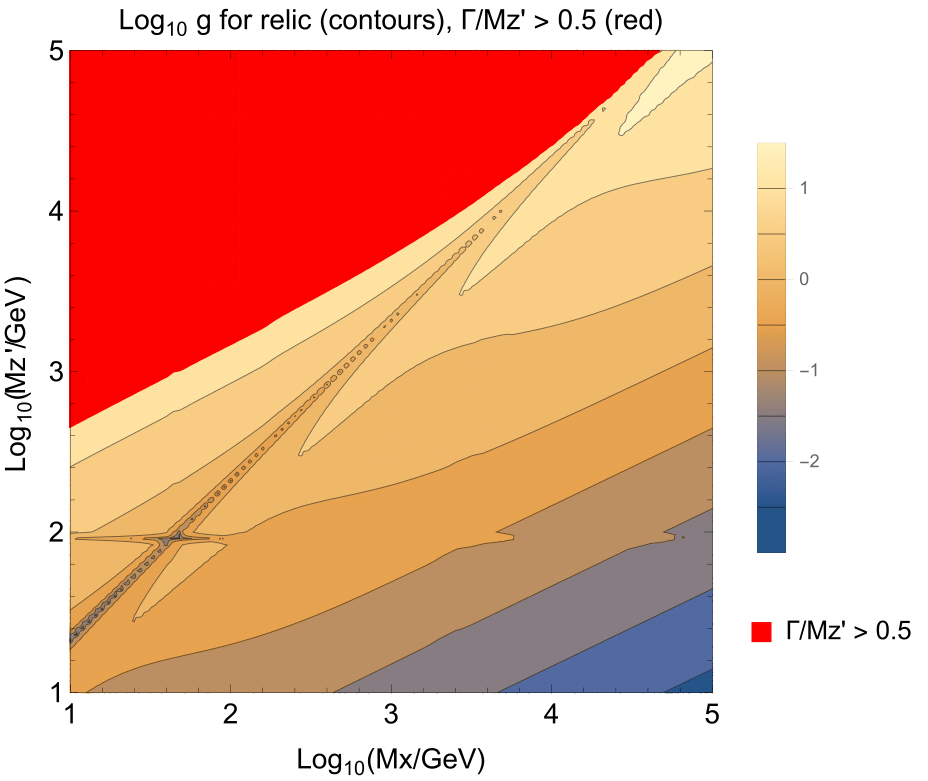}
\includegraphics[width = 0.5\textwidth]{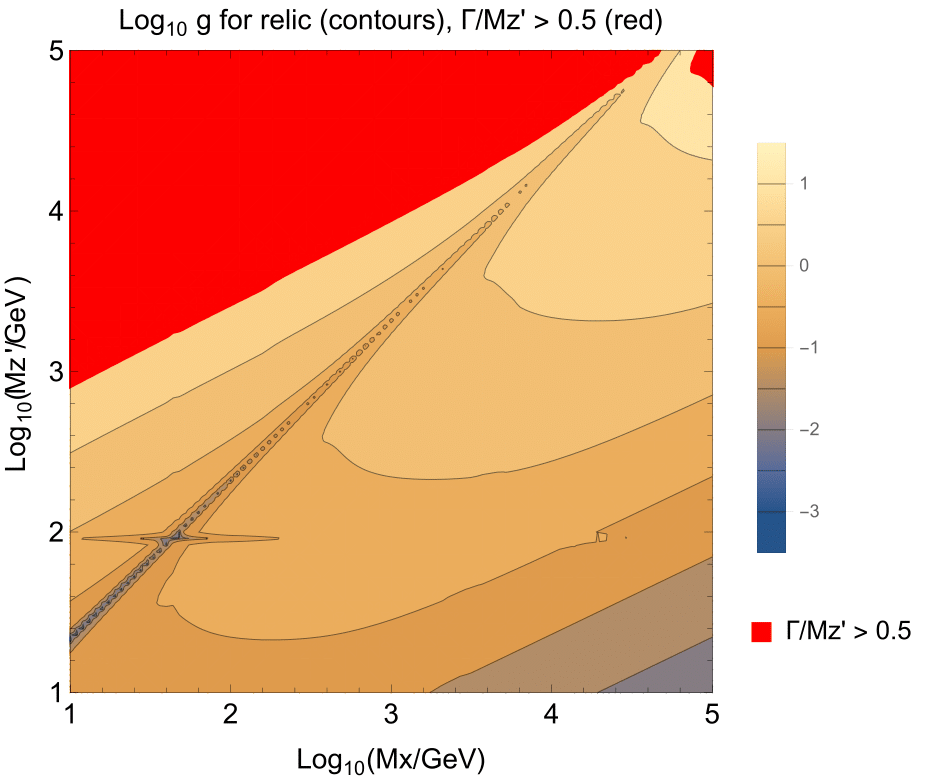}
\caption{\it The $(m_\chi, M_{Z^\prime})$ planes in the leptophobic U(1)$^\prime$
models Eq.~(\protect\ref{Charges1}) (left panel) and Eq.~(\protect\ref{Charges2}) (right panel),
with the value of the gauge coupling $g$ varying across the planes so as to enforce
$\Omega_\chi = \Omega_{CDM}$, as indicated by the colours and solid contours.}
\label{fig:leptophobicvaryg}
\end{figure}

This effect is also visible in Fig.~\ref{fig:leptophobicvaryg}, where the
gauge coupling $g$ is allowed to vary across the $(m_\chi, M_{Z^\prime})$ planes
so as to maintain $\Omega_\chi = \Omega_{CDM}$ in the leptophobic U(1)$^\prime$
models with the quark charges (\ref{Charges1}) (left panel) and (\ref{Charges2}) (right panel).
In the red shaded regions $\Gamma_{Z'}/M_{Z'} > 0.5$. so that the narrow-width approximation
breaks down.

\begin{figure}
\centering
\includegraphics[width = 0.45\textwidth]{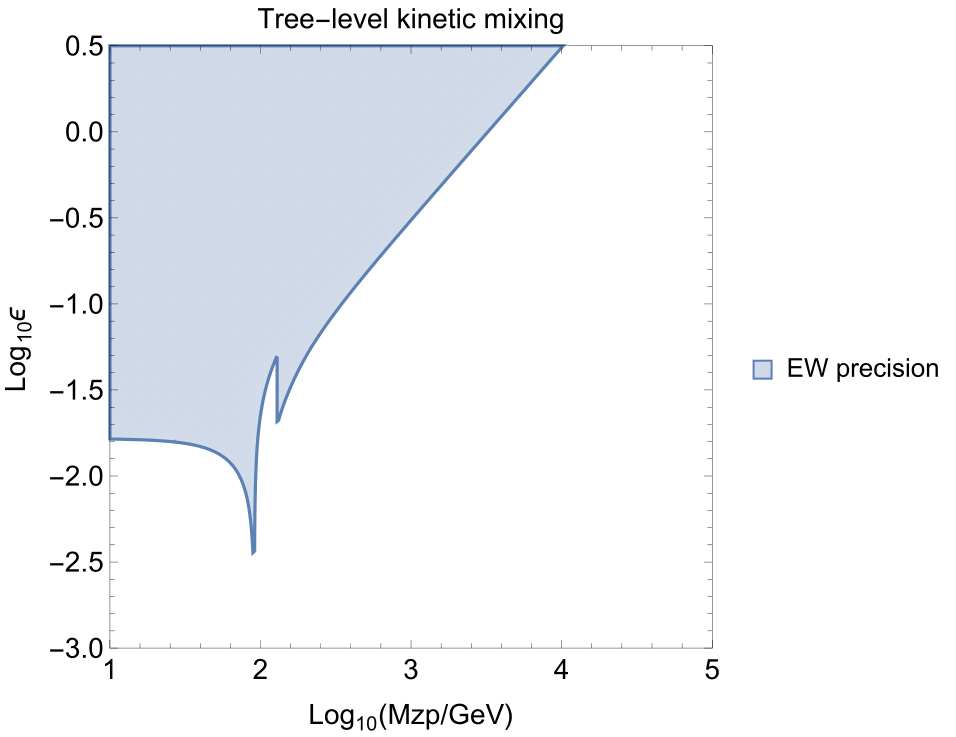} \\
\includegraphics[width = 0.45\textwidth]{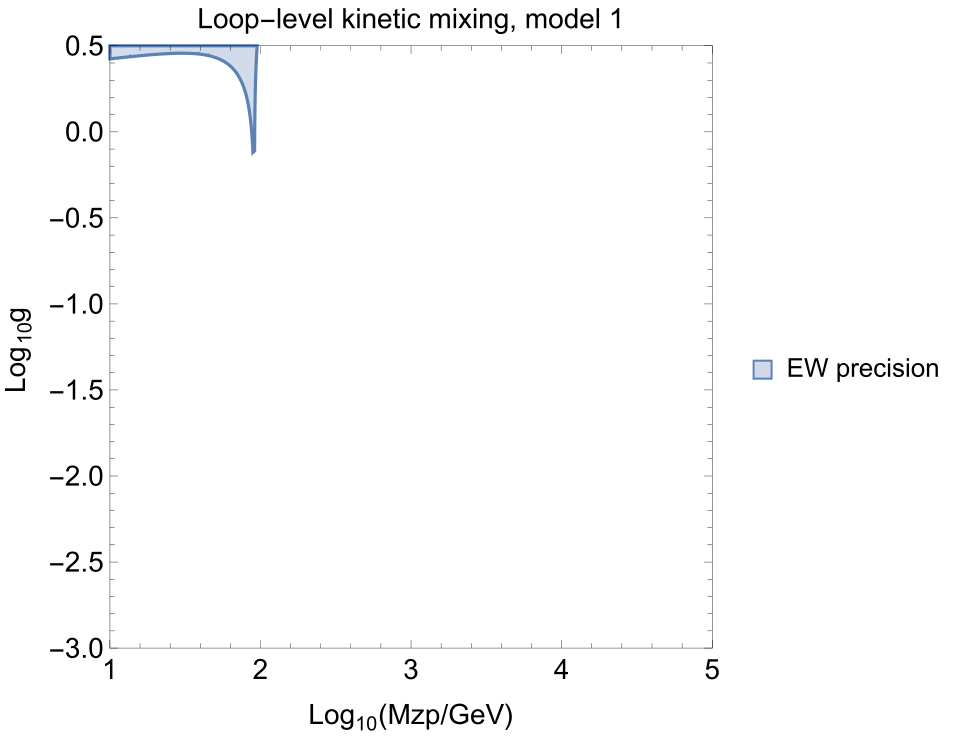}
\includegraphics[width = 0.45\textwidth]{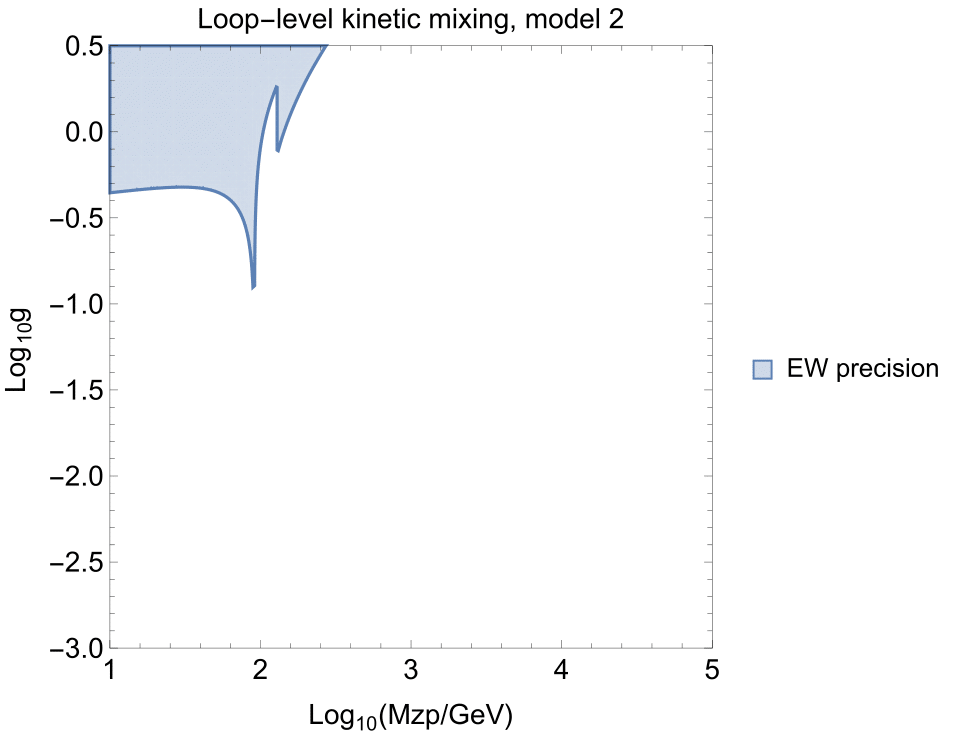}
\caption{\it Upper panel: The constraint on kinetic mixing $\epsilon$ as a function of $M_{Z^\prime}$
imposed by precision measurements of the oblique parameters $S$ and $T$ (and $\rho$).
Lower left panel: The kinetic mixing constraint in the $(M_{Z^\prime}, g)$ plane in the
first leptophobic model (Eq.~(\protect\ref{Charges1})), taking account of
the logarithmic variation of $\epsilon$ and assuming that it vanishes at a renormalization scale of $100$~TeV.
Right panel: The corresponding kinetic mixing constraint in the $(M_{Z^\prime}, g)$ plane in the
second leptophobic model (Eq.~(\protect\ref{Charges2})).}
\label{fig:LoopKineticMixing}
\end{figure}

The upper panel of Fig.~\ref{fig:LoopKineticMixing}
displays the constraint imposed by precision measurements of the oblique
parameters $S, T$ in the $(M_{Z^\prime}, \epsilon)$ plane~\cite{sandt2018}, where $\epsilon$ is the
magnitude of (tree-level) kinetic mixing. For our models however, we will assume that at tree level $\epsilon = 0$, but we cannot avoid generating it at loop-level. In the lower left panel of Fig.~\ref{fig:LoopKineticMixing}, 
we show the constraint in the $(M_{Z^\prime}, g)$ plane that is imposed by the oblique
parameters $S, T$ in the first leptophobic model with $Y^\prime_q = -1/27$ (Eq.~(\ref{Charges1})),
and in the lower right panel the corresponding constraint in the second leptophobic model with $Y^\prime_q = -2/9$ 
(Eq.~(\ref{Charges2})), assuming in both cases that the loop-induced mixing vanishes at the scale of 100~TeV~\footnote{For
consistency, the scale at which the mixing vanishes should not lie within the range of
$M_{Z^\prime}$ displayed in the figures. By choosing the mixing to vanish at the boundary
of the displayed range of $M_{Z^\prime}$, we are applying it in the most conservative
possible way.}. For further details on the electroweak precision constraints, see Appendix~\ref{AppMix}.

This constraint is much weaker than the mass mixing constraint in 
the $Y^\prime$-sequential models that was shown in Fig.~\ref{fig:EWPrecision}, due to both the loop-suppression and the small quark charges present in both models. In
particular, in the case of the second leptophobic model we see in the lower right panel of 
Fig.~\ref{fig:LoopKineticMixing} that for $M_{Z^\prime} < M_Z$ only $g \gtrsim 0.5$ is disallowed,
and that any value of $g < 1$ is allowed for $\log_{10} (M_{Z^\prime} / {\rm GeV}) \gtrsim 2.1$.
Since the first leptophobic model has a smaller quark charge, namely $Y^\prime_q = -1/27$ (Eq.~(\ref{Charges1})),
the constraint on $g$ for any fixed value of $M_{Z^\prime}$ is weaker by a factor 6,
and hence of even less importance, as seen in the lower left panel of Fig.~\ref{fig:LoopKineticMixing}~\footnote{The
glitches seen in the upper and lower right panels of Fig.~\ref{fig:LoopKineticMixing} arise from a mismatch between
our treatments of the precision electroweak constraints using $S$ and $T$ at large $M_{Z'}$ and the $\rho$
parameter at smaller $M_{Z'}$, and are without consequence for our global analysis.}.

\begin{figure}
\centering
\hspace{-2cm}
\includegraphics[width = 0.6\textwidth]{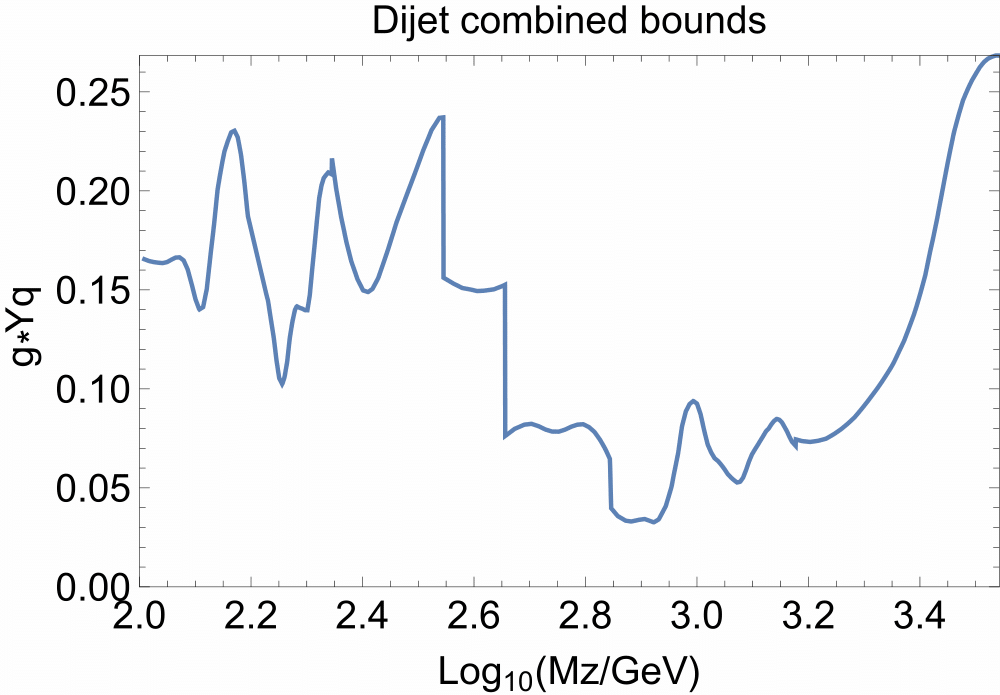} \\
\vspace{0.5cm}
\caption{\it The upper limits on the on the quark coupling
$g \times Y'_q$ obtained from the LHC 13-TeV dijet searches listed in the text.
}
\label{fig:dijet}
\end{figure}

We consider next the dijet bounds on these leptophobic models, which are shown
in Fig.~\ref{fig:dijet}. This shows the constraints on the quark coupling
$g \times Y'_q$ when $m_\chi > M_{Z^\prime}/2$, so that the invisible width vanishes. 
The irregularities in the limit contour arise because several different
13-TeV experimental analyses are combined: 
\begin{itemize}
\item An ATLAS search for resonances 
decaying into boosted quark pairs + a $\gamma$ or a jet
with 36.1/fb for $M_{Z'} < 220$~GeV~\cite{Aaboud:2018zba}, 

\item An ATLAS search for dijets + an ISR $\gamma$
with 15.5/fb for $220$~GeV$ < M_{Z'} < 350$~GeV~\cite{ATLAS:2016bvn}, 
 
\item An ATLAS search for dijets + an ISR jet
with 15.5/fb for $350$~GeV$ < M_{Z'} < 450$~GeV~\cite{ATLAS:2016bvn}, 

\item An ATLAS dijet search
with 3.6 to 29.7/fb for $450$~GeV$ < M_{Z'} < 1500$~GeV~\cite{Aaboud:2018fzt}, 

\item An ATLAS dijet search
with 37.0/fb for $1.5$~TeV$ < M_{Z'}< 3.5$~TeV~\cite{Aaboud:2017yvp}.

\end{itemize}

\noindent

We have also explored the constraints on the leptophobic models coming from monojet
searches at the LHC. To this end, we have modelled the published results from ATLAS~\cite{Aaboud:2017phn} using a rapid
recasting procedure that reproduces the published experimental results within the
quoted $\pm 1 \sigma$ uncertainty, the main deviations being associated with binning
effects in the experimental analysis and theoretical modelling. Practical details are described in Appendix~B.

\begin{figure}
\centering
\includegraphics[width = 0.45\textwidth]{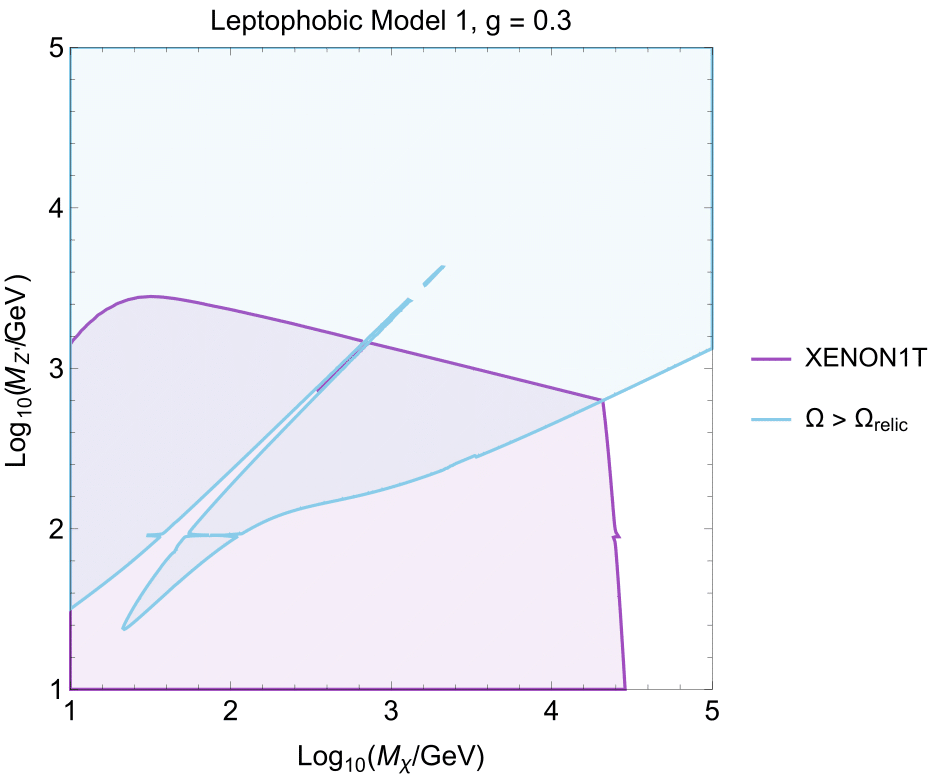}
\includegraphics[width = 0.45\textwidth]{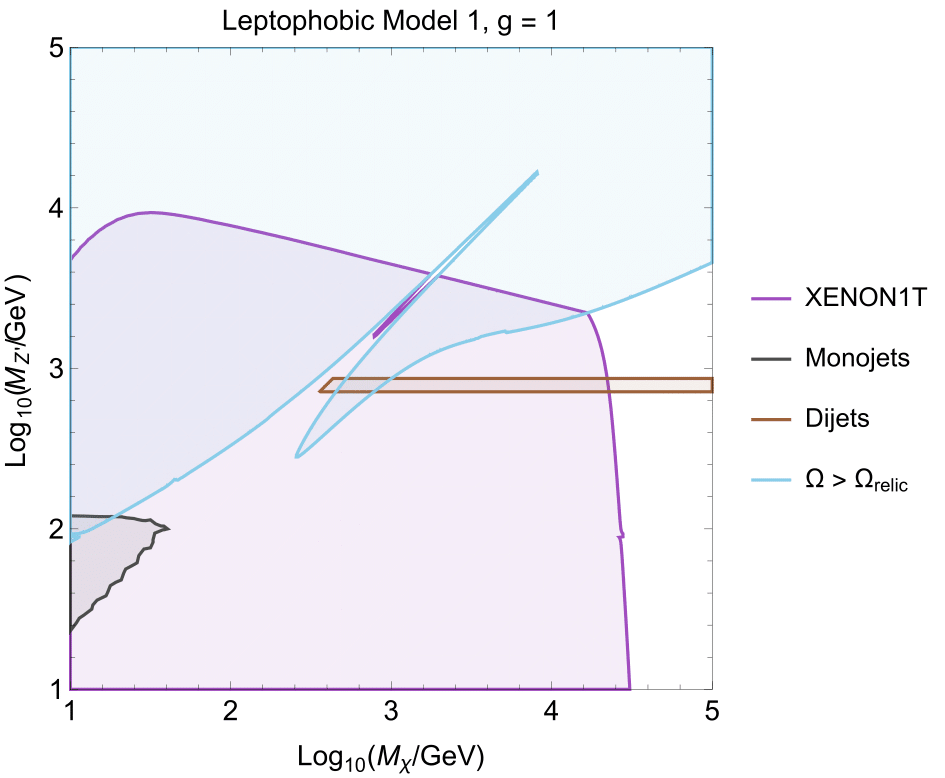}\\
\includegraphics[width = 0.45\textwidth]{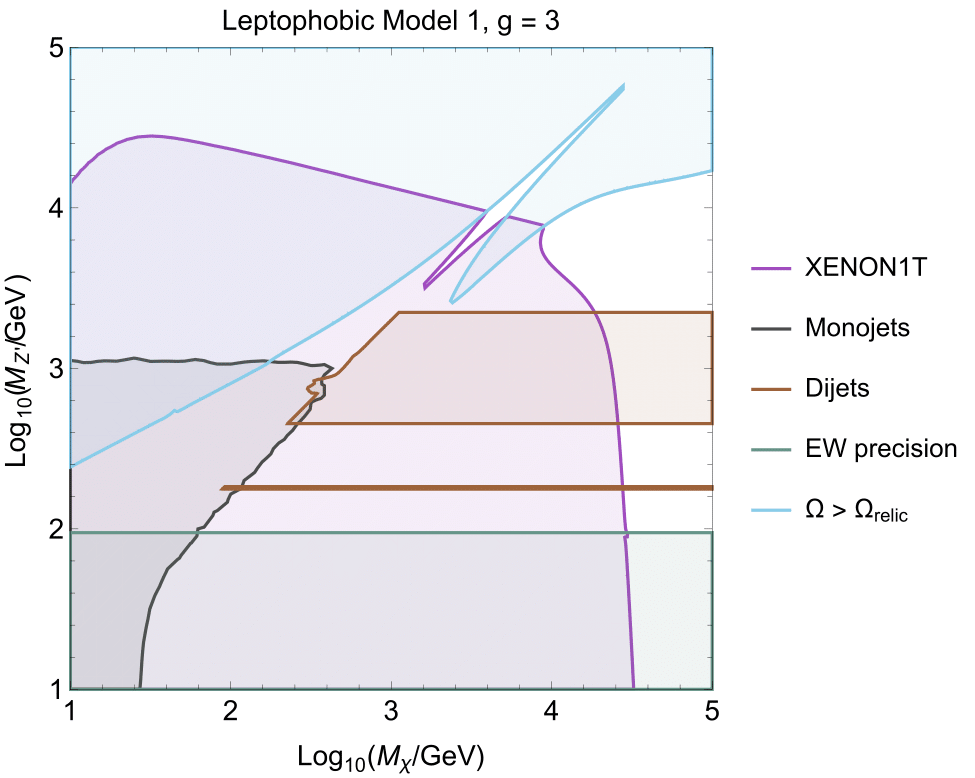}
\caption{\it The $(m_\chi, M_{Z^\prime})$ planes for leptophobic model 1, for a gauge coupling $g = 0.3$ (upper left),
$g = 1.0$ (upper right) and $g = 3.0$ (lower). The
solid blue lines are the contours where $\Omega_\chi = \Omega_{CDM}$, and $\Omega_\chi > \Omega_{CDM}$
in the regions shaded blue. The dark grey band is excluded by the most recent ATLAS monojet search and the bands shaded brown are excluded by ATLAS dijet searches. The regions shaded purple
are excluded by direct searches for dark matter scattering, and the regions shaded green are excluded by precision electroweak data.}
\label{fig:Lept1Fixedg}
\end{figure}
\begin{figure}
\centering
\includegraphics[width = 0.45\textwidth]{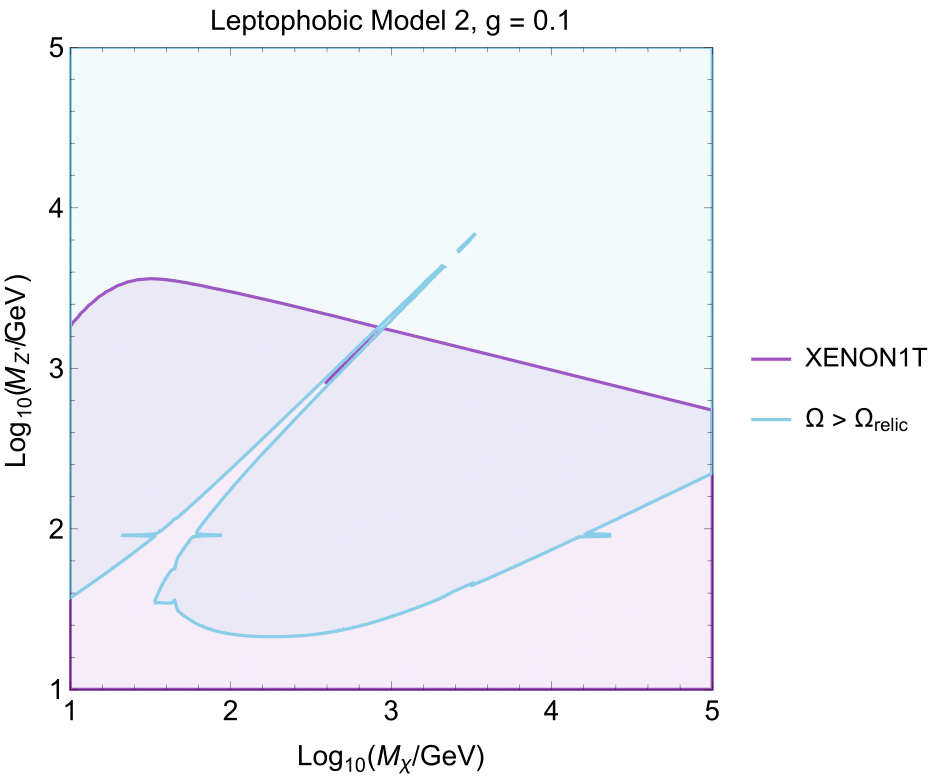}
\includegraphics[width = 0.45\textwidth]{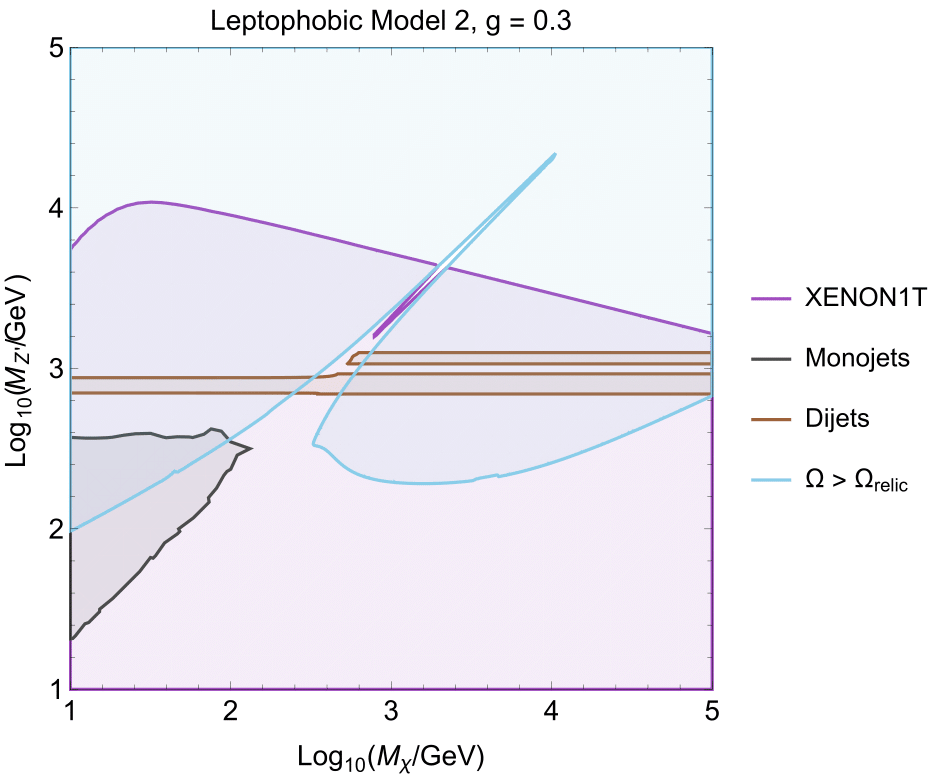}\\
\includegraphics[width = 0.45\textwidth]{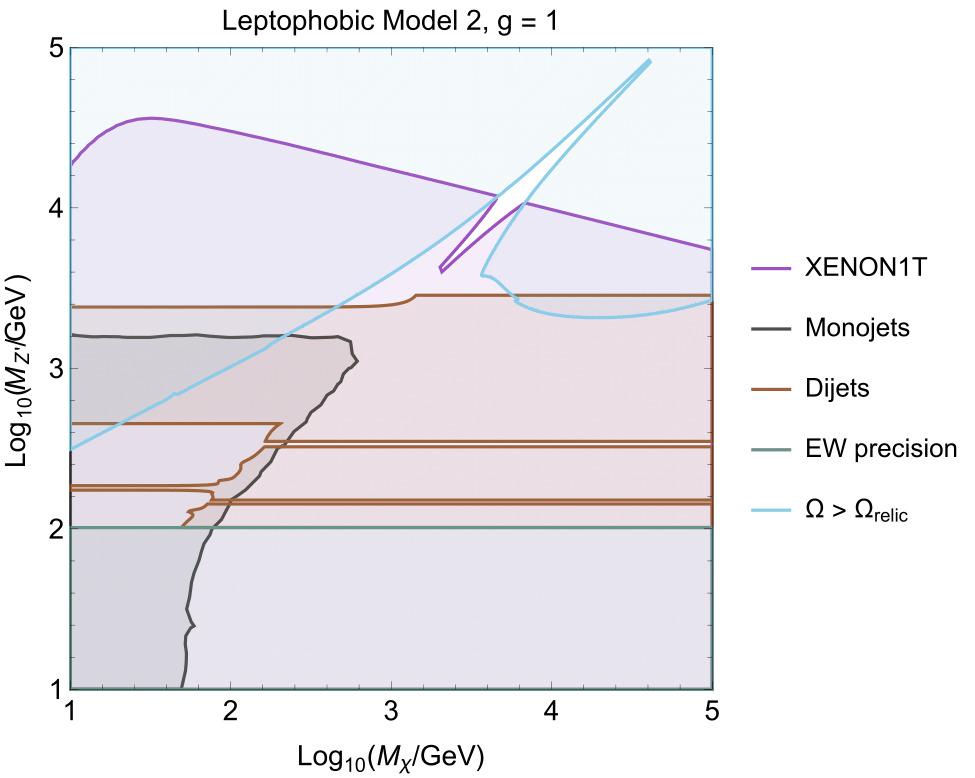}
\caption{\it The $(m_\chi, M_{Z^\prime})$ planes for leptophobic model 2, for a gauge coupling $g = 0.1$ (upper left),
$g = 0.3$ (upper right) and $g = 1.0$ (lower). The
solid blue lines are the contours where $\Omega_\chi = \Omega_{CDM}$, and $\Omega_\chi > \Omega_{CDM}$
in the regions shaded blue. The dark grey band is excluded by the most recent ATLAS monojet search and the bands shaded brown are excluded by ATLAS dijet searches. The regions shaded purple
are excluded by direct searches for dark matter scattering, and the regions shaded green are excluded by precision electroweak data.
}
\label{fig:Lept2Fixedg}
\end{figure}

Next we show summary plots of all relevant constraints for fixed gauge couplings, in which we treat the relic density as an upper limit rather than a strict requirement, for leptophobic model 1 in Fig.~\ref{fig:Lept1Fixedg} and for leptophobic model 2 in Fig.~\ref{fig:Lept2Fixedg}. We see 
that relic density considerations along with the direct detection constraint rule out much of the parameter space, with LHC searches being less important. 
In particular, the monojet constraint is unimportant for $g = 0.3$, but makes an appearance for $g = 1$ and becomes more important
for $g = 3$. At low $m_\chi$ and $M_{Z'}$ the monojet signal would fall into the low $E_{T,\text{miss}}$ selection, whereas at higher masses
the signal is best constrained by the higher $E_{T,\text{miss}}$ selection. The band structure of the region excluded by the LHC dijet searches
arises because of the irregularity in the combined cosntraint seen in Fig.~\ref{fig:dijet}.

\begin{figure}
\centering
\includegraphics[width = 0.45\textwidth]{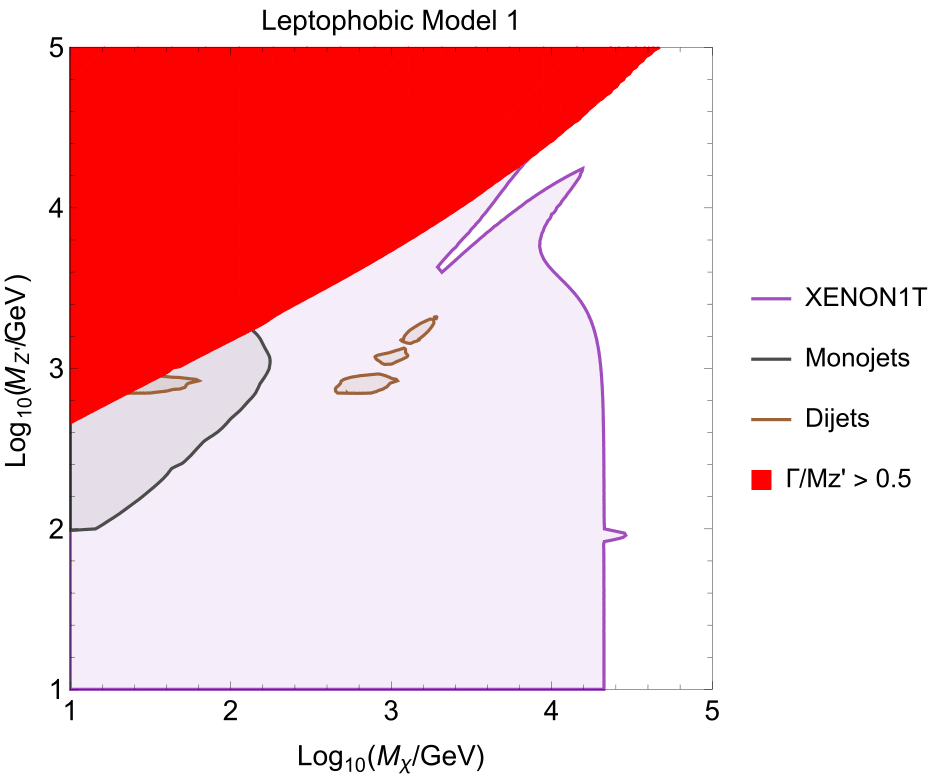}
\includegraphics[width = 0.45\textwidth]{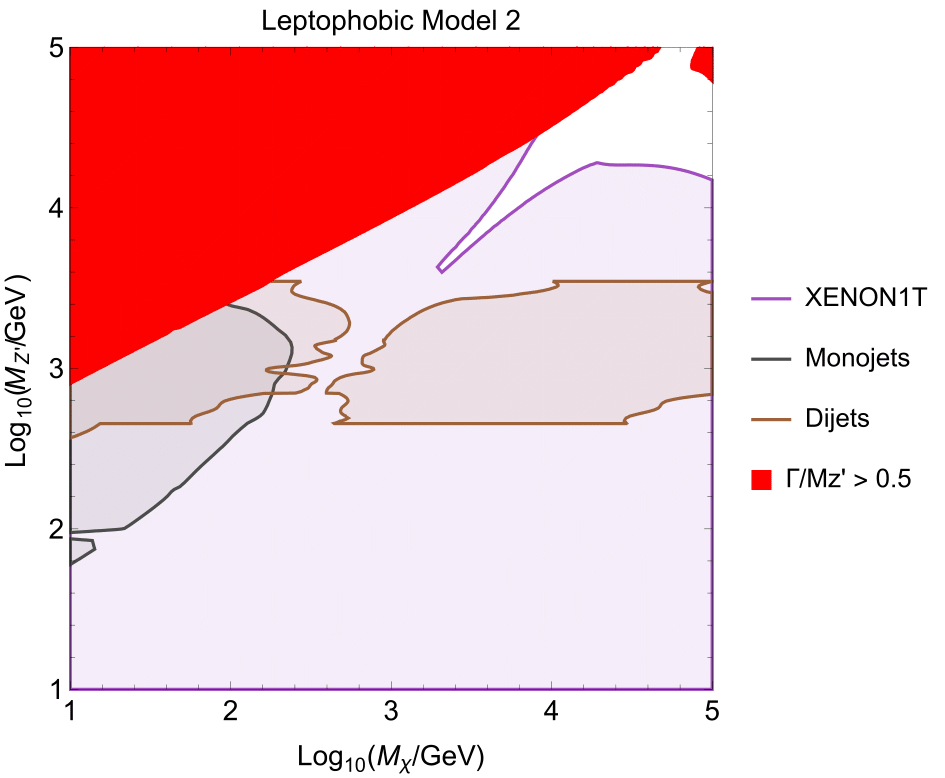}
\caption{\it Left panel: Compilation of constraints in the $(m_\chi, M_{Z^\prime})$ plane in the
first leptophobic model (Eq.~(\protect\ref{Charges1})), here coupling is varied to obtain good relic abundance. Right panel: The corresponding
compilation of constraints in the $(m_\chi, M_{Z^\prime})$ plane in the
second leptophobic model (Eq.~(\protect\ref{Charges2})).}
\label{fig:LeptophobicCompilations}
\end{figure}

Finally, we show in the left and right panels of Fig.~\ref{fig:LeptophobicCompilations}, respectively,
compilations of the various phenomenological constraints in the $(m_\chi, M_{Z^\prime})$ planes
for the first and second leptophobic models (Eqs.~(\ref{Charges1}, \ref{Charges2})), varying $g$ so as to
obtain the correct total cold dark matter density. The monojet constraints 
(black lines and grey shading) are quite similar
in the two models, despite the differences in their $Z'$-quark couplings~\footnote{This is because the 
gauge coupling determined via the relic density for model 1 is higher than that for model 2 because it has a smaller quark charge, compensating for the smaller quark charge that enters the monojet production cross-section. In addition, model 1 has a higher invisible branching fraction.},
and limited to $\log_{10} (M_{Z'}/{\rm GeV}) \lesssim 3.3$ to 3.4. 
We see that the
dijet constraint (orange lines and shading) is generally weaker in the first model, as was to be expected in view of its
smaller $Z'$-quark couplings.
We also see that in both cases the direct DM detection constraints (purple lines and shading) are stronger than those from
the dijet and monojet constraints. In the first leptophobic model, when $\log_{10} (m_\chi/{\rm GeV}) \lesssim 4$ the direct DM
scattering constraint enforces $\log_{10} (M_{Z'}/{\rm GeV}) \gtrsim 3.2$, which is attained along
the diagonal line where $M_{Z'} = 2 m_\chi$ and rapid resonant annihilation requires a smaller
value of the coupling $g$, reducing the scattering cross section. In the second leptophobic model the direct dark 
matter constraint imposes $\log_{10} (M_{Z'}/{\rm GeV}) \gtrsim 3.2$ in all the plane displayed.

The importance of the direct DM scattering constraint in Fig.~\ref{fig:LeptophobicCompilations} arises from the vector nature of the 
coupling of the DM particle to quarks in the two minimal leptophobic models (Eqs.~(\ref{Charges1}, \ref{Charges2}))
proposed in~\cite{usEFT1}. It would be possible, in principle, to construct non-minimal leptophobic models
in which the quark couplings are axial, in which case the impact of the direct DM scattering constraint would be
reduced.
In this hypothetical case indirect constraints on DM annihilations, e.g., from searches for $\chi \chi \to \gamma$ + X
in dwarf spheroidal galaxies~\cite{dwarves}, would play a role for $m_\chi \lesssim 50$~GeV. 
However, we do not consider this case any further, and away from resonance these indirect searches
play no role in constraining our benchmark vector-like leptophobic models. 

\begin{figure}
\centering
\includegraphics[width = 0.45\textwidth]{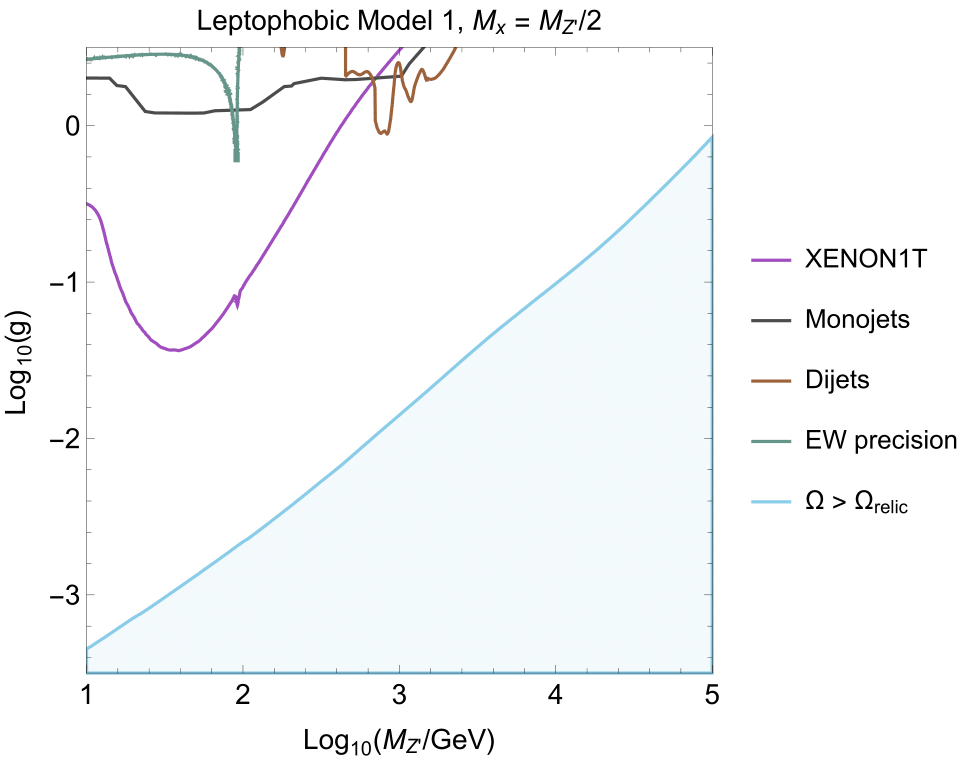}
\includegraphics[width = 0.45\textwidth]{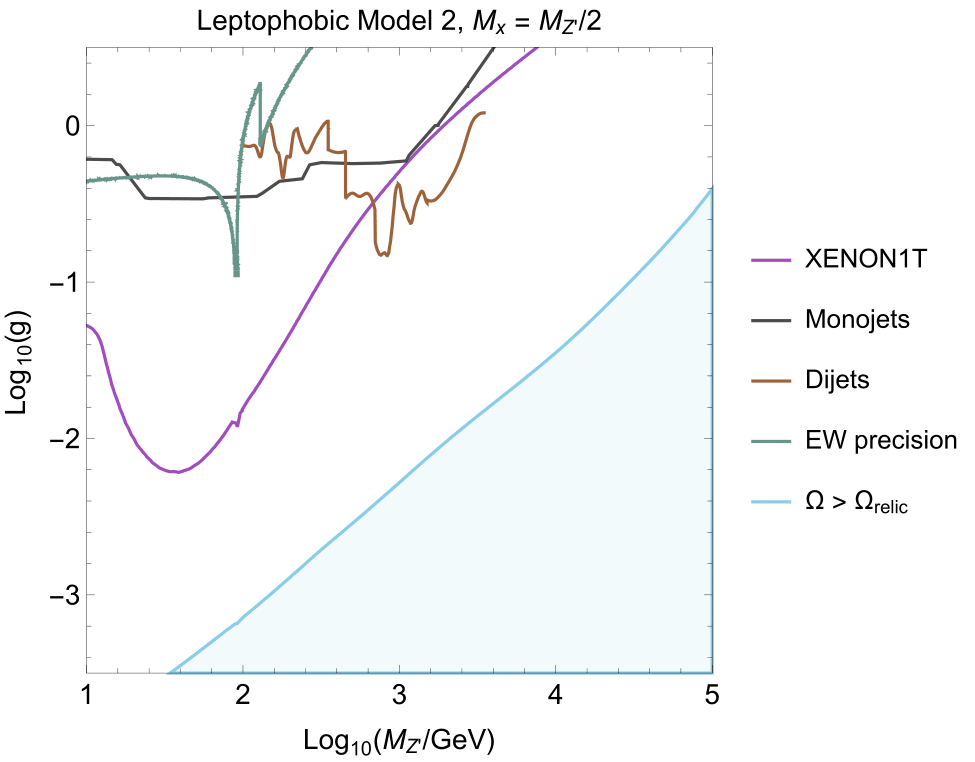}
\caption{\it The $(M_{Z^\prime},g)$ planes for leptophobic model 1 (left) and 2 (right), with $m_\chi = M_{Z^\prime}/2$ for resonant annihilation. The
solid blue lines are the contours where $\Omega_\chi = \Omega_{CDM}$, and $\Omega_\chi > \Omega_{CDM}$
in the regions shaded blue. The region above the dark grey line is excluded by the most recent ATLAS monojet search and the region above the line shaded brown is excluded by ATLAS dijet searches. The region above the purple line
is excluded by direct searches for dark matter scattering, and the region above the green line is excluded by precision electroweak data.}
\label{fig:LeptRes}
\end{figure}

As in the previous leptophilic models, in narrow strips of resonant annihilation near 
the $Z'$ peak where $m_\chi \simeq M_{Z^\prime}/2$, the gauge coupling $g$ may
be significantly smaller while also reproducing the observed relic density. To investigate to what extent, if at all, the other 
experimental constraints can exclude this region, we show the $(M_{Z^\prime},g)$ plane for both leptophobic models in Fig.~\ref{fig:LeptRes}. 
We see that, in both cases, the correct total cold dark matter density can be obtained for any value of $M_{Z'}$
without coming into conflict with data from direct detection, the LHC and electroweak precision data. As in the case of the vector-like
leptophobic model, this feature is too narrow to be visible in the $(m_\chi, M_{Z'})$ planes shown in Fig.~\ref{fig:LeptophobicCompilations}.

\section{Discussion and Conclusions\label{sec:conclusions}}

We have studied four benchmark models of dark matter taken from~\cite{usEFT1}, whose interactions are mediated by an anomaly-free 
$Z'$ boson. Two of these models are leptophilic, one with a vector-like coupling of the dark matter particle
to the $Z'$, and one with an axial coupling. The other two models are leptophobic, with the gauge anomalies cancelled by different sets of additional particles
in the dark sector. We have considered the phenomenological constraints coming from the overall density of cold dark matter,
direct searches for dark matter scattering, from LHC searches for dileptons, dijets and monojets, and from precision electroweak
measurements.

We have found that the vector-like leptophilic model is extremely tightly constrained by both dilepton constraints,
and especially modifications to the $S$ and $T$ electroweak parameters, 
which rule out almost completely the areas of parameter space where we obtain good relic abundance.  
There is, however, a very small region of parameter space still available where both $m_\chi$ and 
$M_{Z'}$ have masses of several TeV.  This region may be accessible to improvements in future constraints on 
electroweak precision variables before future enhancements in direct detection constraints. 
In addition to this region, there is a continuous line of solutions constrained to a very narrow allowed strip where $m_\chi \simeq M_{Z'}/2$.  

The axial leptophilic model is excluded for $M_{Z'} < 10$~TeV completely, again by dilepton constraints and modifications to the electroweak variables. Therefore,
this model requires modifications if it is to survive in the energy window that we are considering, namely that of interest to the LHC.

The two leptophobic models both have larger allowed regions where $\log_{10} (m_\chi/ {\rm GeV}) \gtrsim 3.2$,
as well as narrow allowed strips where $m_\chi \simeq M_{Z'}/2$.  The interesting regions of these models are 
generally safe in terms of their effect upon electroweak precision variables, as well as evading
the dilepton bounds that constrain tightly the previous models. The monojet constraints on both models are relatively weak compared to the other constraints. The leptophobic model with a triplet
of `dark' particles has a stronger $Z'$ coupling to quarks, so that the dijet searches are stronger.
However, despite this, the constraint from direct detection limits is the strongest constraint on the parameter spaces of both leptophobic models.
Since the LHC centre-of-mass energy will not be increased substantially, whereas the integrated luminosity will increase
by almost two orders of magnitude compared to that analyzed so far,
we expect that the improvement in dijet constraints will be mainly in terms of coupling rather than $Z'$ mass.  We
therefore expect future direct dark matter detection experiments to continue to impose stronger constraints than future collider results.

We have shown that $Z'$ models similar to the spin-one simplified models widely studied in the literature are either very strongly constrained (the Y-sequential models) or must feature exotic fermions charged under the SM gauge group (including SU(2) multiplets). In the latter case, it would be interesting in the future to study novel experimental constraints that might arise from the presence of such exotic fermions. On the theoretical side, it would be of interest to come up with an anomaly-free theory that features a purely axial coupling to dark matter, since this would allow a greater deal of complementarity between LHC and direct detection constraints. For our benchmark models, we have found that complementarity between different experimental constraints is not so simple to achieve.

The great progress made in recent years in exploring new physics scenarios at colliders and in underground experiments still leaves uncovered regions of parameter space which will be probed by the next generation of colliders. In particular we have shown how dijet and dilepton searches can set the strongest constraints when the DM annihilation is on the $Z'$ resonance. We look forward to the continued exploration of simplified anomaly-free models of dark matter from both the theory community and future experimental data.

\section*{Acknowledgements}

The work of JE and MF was supported partly by the STFC Grant ST/P000258/1. In addition, MF and PT are funded by the European Research Council under the European Union's Horizon 2020 programme (ERC Grant Agreement no.648680 DARKHORIZONS).

\appendix

\section{$Z-Z'$ mixing\label{AppMix}}

We follow the approach in \cite{babumix,Kahlhoefer:2015bea,duerrmix}, assuming a Lagrangian with both mass and kinetic mixing:
\begin{align}
  {\cal L} =& \; {\cal L}_\text{SM}
  -\frac{1}{4}{\hat{F}}^{\prime \mu\nu}\hat{F}'_{\mu\nu} + {\frac{1}{2}} m_{\hat
    Z'}^2 \hat{X}_\mu \hat{X}^\mu - {\frac{1}{2}} \sin \epsilon\, \hat{B}_{\mu\nu} {\hat{F}}^{\prime\mu\nu} +\delta
  m^2 \hat{Z}_\mu \hat{X}^\mu 
\end{align}
where $\hat{Z}\equiv \cos\hat\theta_{W} \hat{W}^3- \sin\hat\theta_W \hat{B}$ and ${\hat{F}}^{\prime\mu \nu} \equiv \partial^\mu \hat{X}^\nu - \partial^\nu \hat{X}^\mu$.

The Lagrangian can be transformed to the mass basis, with canonical kinetic terms, via the following transformations:
\begin{align}
\left(\begin{array}{c} \hat B_\mu \\ \hat W_\mu^3 \\ 
\hat{X}_\mu \end{array}\right) & = \left(\begin{array}{ccc} 1 & 0 &
    -\tan\epsilon \\ 0 & 1 & 0 \\ 0 & 0 &
    1/\cos\epsilon \end{array}\right)
\left(\begin{array}{c} B_\mu \\ W_\mu^3 \\ X_\mu \end{array}\right) 
\end{align}
\begin{align}
\left(\begin{array}{c} B_\mu \\ W_\mu^3 \\ X_\mu \end{array}\right) &
= \left(\begin{array}{ccc}
    \cos\hat\theta_{W} & -\sin\hat\theta_{W} \cos\xi &  \sin\hat\theta_{W} \sin\xi \\
    \sin\hat\theta_{W} & \cos\hat\theta_{W} \cos\xi & - \cos\hat\theta_{W} \sin\xi \\
    0 & \sin\xi & \cos\xi
\end{array} \right) 
\left(\begin{array}{c} A_\mu \\ Z_\mu \\ Z'_\mu \end{array}\right)
 \;
\end{align}
where we identify $A$, $Z$ and $Z^{\prime}$ as the physical fields, with $\xi$ determined by
\begin{align}
  \tan(2\xi)=\frac{-2\cos\epsilon(\delta m^2+m_{\hat Z}^2 \sin\hat\theta_{W} \sin\epsilon)} {m_{\hat Z'}^2-m_{\hat
      Z}^2 \cos\epsilon^2 +m_{\hat Z}^2\sin\hat\theta_{W}^2 \sin\epsilon^2
    +2\,\delta m^2\,\sin\hat\theta_{W} \sin\epsilon} 
\end{align}
The impact on electroweak precision observables can then be calculated using the $S$ and $T$ parameters~\footnote{To 
lowest order in $\xi$, $\cos\hat\theta_{W} = \cos\theta_{W}$ and $\sin\hat\theta_{W} = \sin\theta_{W}$.}:
\begin{align}
  \alpha \, S = & 4 \, \cos\theta_{W}^2 \, \sin\theta_{W} \, \xi \, (\epsilon - \sin\theta_{W} \, \xi)  \\
  \alpha \, T = & \xi^2 \left(\frac{m_{Z'}^2}{m_{Z}^2}-2\right) + 2 \, \sin\theta_{W} \, \xi \, \epsilon 
\end{align}
where $\cos\theta_{W}$ is the cosine of the electroweak mixing angle and $\alpha=e^2/4\pi$ is the electroweak coupling. 
We use for the numerical values of the $S$ and $T$ parameters the recent fit~\cite{sandt2018}.
However, for smaller $Z'$ masses it is more suitable to use the $\rho$ parameter
\begin{equation}
 \rho - 1 = \frac{\cos^2\theta_{W} \, \xi^2}{\cos^2\theta_{W} - \sin^2\theta_{W}} \left(\frac{m_{Z'}^2}{m_{Z}^2}-1\right) 
\end{equation}
and we use the $\rho$ parameter instead of the $S$ and $T$ parameters when $M_{Z^{\prime}}< \sqrt{2} M_Z$. The reader who is paying attention may notice some glitches in some diagrams at those places in parameter space where we switch from using the $S,T$ variables to $\rho$.

For the Y-sequential models in Section 2 we neglect kinetic mixing,
since the effect of mass mixing is much stronger. We then find
\begin{align}
\delta m^2  = \frac{1}{2}\frac{e g \, Y^{\prime}_H}{\sin\theta_{W} \cos\theta_{W}}v^2  
\end{align}
where $g$ is the $U(1)^{\prime}$ gauge coupling, $Y^{\prime}_H$ is the Higgs charge under $U(1)^{\prime}$,
and $v$ is the SM Higgs vev.

For the leptophobic models, $Y^{\prime}_H = 0$ so there is no mass mixing effect, and
we assume also that tree-level kinetic mixing vanishes. However, it is unavoidably generated at loop level. 
Conservatively, we assume that $\epsilon = 0$ at $\Lambda = 100$ TeV, such that at a lower scale~\cite{Carone:1995pu}
\begin{equation}
\epsilon(\mu) = \frac{e \, g \, Y^{\prime}_q}{2 \pi^2 \, \cos \theta_\text{W}} \log \frac{\Lambda}{\mu}
\end{equation}
In calculating our constraints we set $\mu = M_{Z'}$.

\section{Monojet recast}

In implementing the LHC monojet constraints, we adopt the rescaling procedure proposed in \cite{Jacques:2015zha},
generating monojet samples across a grid of $Z'$ and dark matter particle masses, 
and then rescaling to other points of parameter space.

\begin{figure}
\centering
\includegraphics[width = 0.45\textwidth]{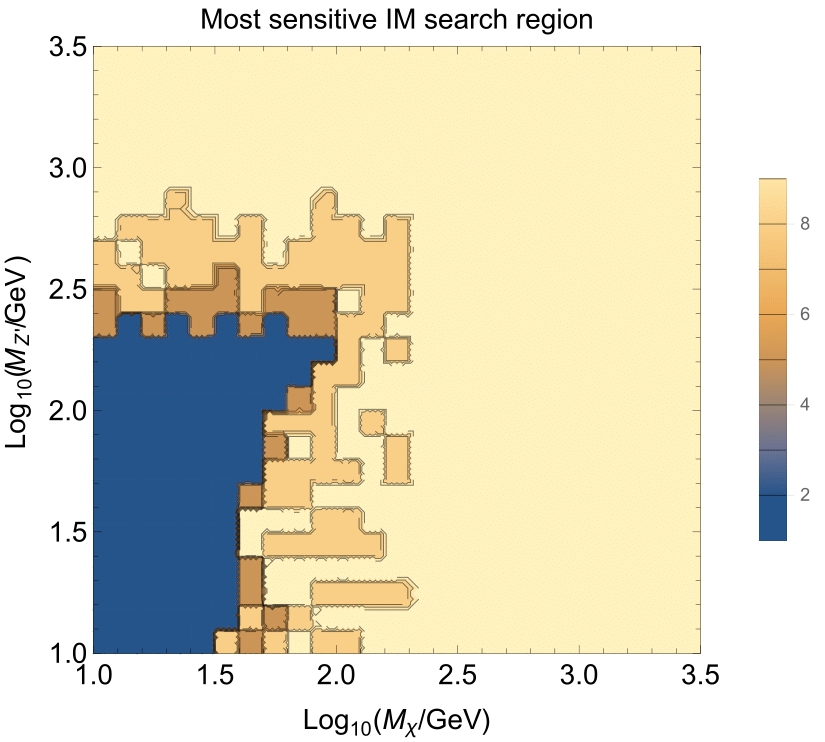}
\includegraphics[width = 0.45\textwidth]{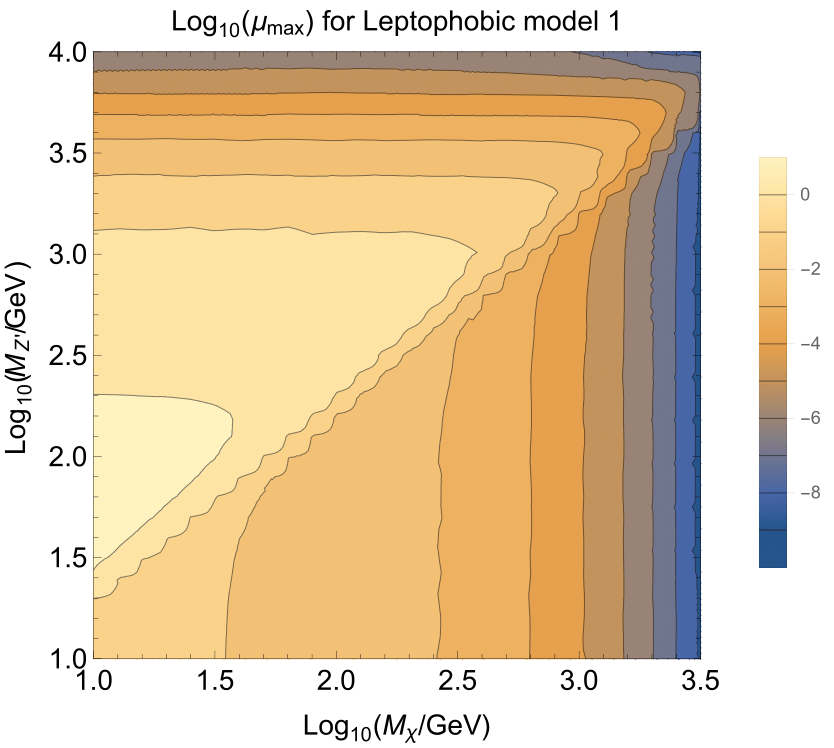}
\caption{\it Left: the most sensitive search region, numbered 1-10 as the inclusive search regions IM1-10 
defined by ATLAS ~\cite{Aaboud:2017phn}. The most sensitive search region is the one that gives the largest 
$\mu$ factor. At low masses, IM1 is the most sensitive, whereas at high masses, IM9 is the most sensitive. Right: Contours of $\mathrm{Log_{10}}\mu$ (see text for definitions) in the most sensitive search region.}
\label{fig:monoBins}
\end{figure}

For the constraints, we use the inclusive selection of the latest ATLAS monojet search with 
36.1$~fb^{-1}$ of integrated luminosity~\cite{Aaboud:2017phn}. We calculate the exclusion 
using each of the missing energy selections defined by ATLAS, IM1 - IM10, 
corresponding to various $E_{T,\mathrm{miss}}$ cuts: $E_{T,\mathrm{miss}} > 250$ GeV for IM1 
up to $E_{T,\mathrm{miss}} > 1000$ GeV for IM10. We calculate $\mu$ as
\begin{equation}
\mu = \frac{\sigma(g = 1, \Gamma = 0.01 M_{Z^\prime})}{\sigma_{95\%}}
\end{equation}
for each separate $E_{T,\mathrm{miss}}$ cut defined in each search region, 
where $\sigma_{95\%}$ is the cross section excluded by ATLAS at the 95\% CL. 
We show which search region is most constraining in the left panel of Fig.~\ref{fig:monoBins}, 
and the corresponding $\mu$ factor in leptophobic model 1 is shown in the right panel of Fig.~\ref{fig:monoBins}.

We then scale this $\mu$ to different points of parameter space by a factor (for fixed charges) 
$g^4/\Gamma$ for the on-shell region, and $g^4$ for the off-shell region, where $\Gamma $ is the width of the Z' boson.

\begin{figure}
\centering
\includegraphics[width = 0.7\textwidth]{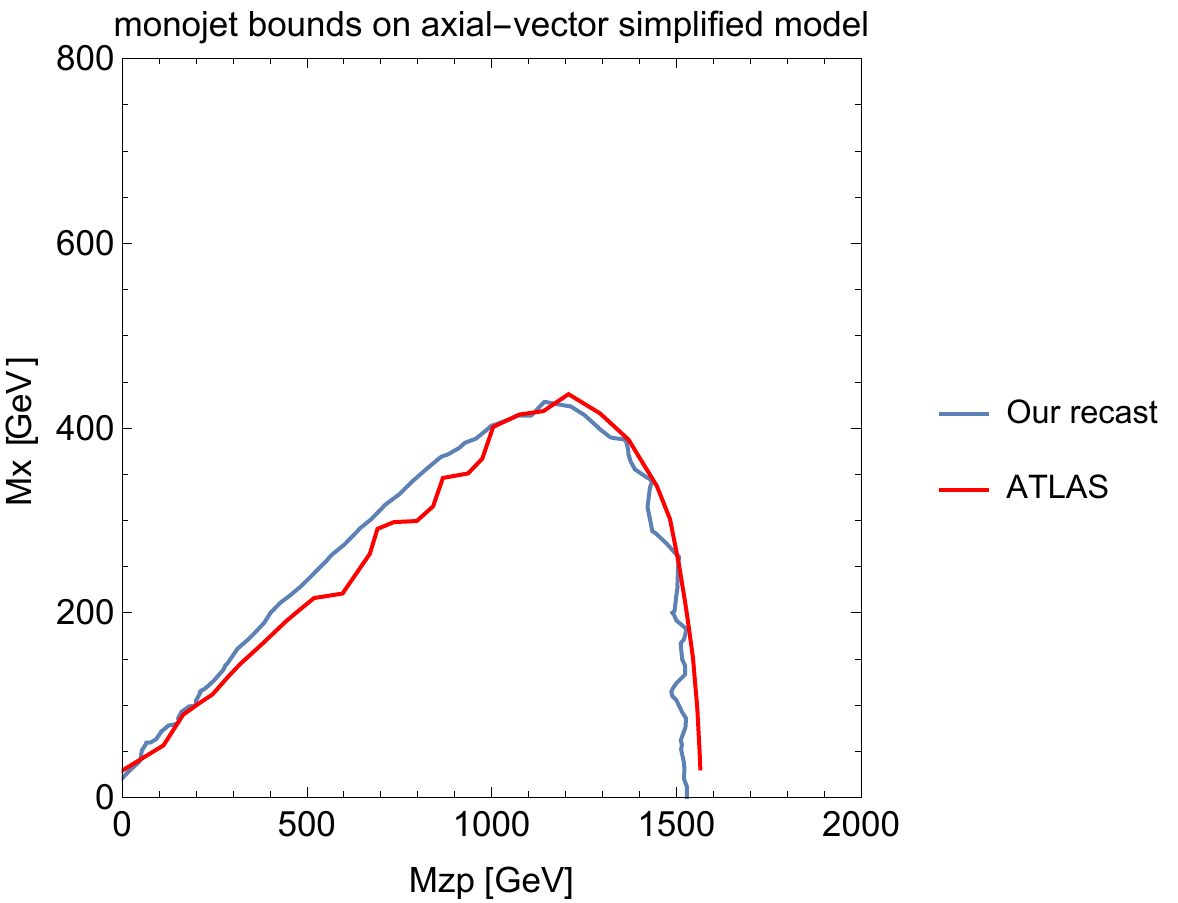}
\caption{\it Parameter points excluded by our recast of the inclusive search (blue) compared to the published results from the exclusive monojet search~\cite{Aaboud:2017phn} (red), 
for the axial-vector simplified model, with fixed couplings of $g_q = 0.25$ and $g_{DM} = 1.0$.}
\label{fig:validation}
\end{figure}

We note that the limit we obtain is approximate, since we do not simulate parton shower or detector effects, and we include the generation of only one hard jet at parton level in {\tt Madgraph}~\cite{madgraph}. However we have validated our approach by reproducing the published results from ATLAS for the axial-vector simplified model, asseen in Fig.~\ref{fig:validation}.

\begin{thebibliography}{99}

\bibitem{Zwicky}
F.~Zwicky,
  Helv.\ Phys.\ Acta {\bf 6} (1933) 110
   [Gen.\ Rel.\ Grav.\  {\bf 41} (2009) 207]
  doi:10.1007/s10714-008-0707-4;
  Astrophys.\ J.\  {\bf 86} (1937) 217
  doi:10.1086/143864.
  
\bibitem{Rubin}
V.~C.~Rubin and W.~K.~Ford, Jr.,
  Astrophys.\ J.\  {\bf 159} (1970) 379
  doi:10.1086/150317;
  V.~C.~Rubin, N.~Thonnard and W.~K.~Ford, Jr.,
  Astrophys.\ J.\  {\bf 238} (1980) 471
  doi:10.1086/158003;
  Y.~Sofue and V.~Rubin,
  Ann.\ Rev.\ Astron.\ Astrophys.\  {\bf 39} (2001) 137
  doi:10.1146/annurev.astro.39.1.137
  [astro-ph/0010594].
  
\bibitem{Peebles}
J.~P.~Ostriker, P.~J.~E.~Peebles and A.~Yahil,
  Astrophys.\ J.\  {\bf 193} (1974) L1
  doi:10.1086/181617.
 

\bibitem{planck}
  P.~A.~R.~Ade {\it et al.} [Planck Collaboration],
  Astron.\ Astrophys.\  {\bf 594} (2016) A13
  doi:10.1051/0004-6361/201525830
  [arXiv:1502.01589 [astro-ph.CO]].

 \bibitem{leeweinberg}
  B.~W.~Lee and S.~Weinberg,
  Phys.\ Rev.\ Lett.\  {\bf 39} (1977) 165.
  doi:10.1103/PhysRevLett.39.165
  
  \bibitem{Hut}
  P.~Hut,
  Phys.\ Lett.\ B {\bf 69} (1977) 85
   [Phys.\ Lett.\  {\bf 69B} (1977) 85].
  doi:10.1016/0370-2693(77)90139-3

\bibitem{hutolive}
  P.~Hut and K.~A.~Olive,
  Phys.\ Lett.\  {\bf 87B} (1979) 144.
  doi:10.1016/0370-2693(79)90039-X
  
\bibitem{LHC4thneutrino}
  P.~Achard {\it et al.} [L3 Collaboration],
  Phys.\ Lett.\ B {\bf 517} (2001) 75
  doi:10.1016/S0370-2693(01)01005-X
  [hep-ex/0107015].

\bibitem{BertoneHooperSilk}
G.~Bertone, D.~Hooper and J.~Silk,
  Phys.\ Rept.\  {\bf 405} (2005) 279
  doi:10.1016/j.physrep.2004.08.031
  [hep-ph/0404175].

\bibitem{EHNOS}
J.~R.~Ellis, J.~S.~Hagelin, D.~V.~Nanopoulos, K.~A.~Olive and M.~Srednicki,
  Nucl.\ Phys.\ B {\bf 238} (1984) 453
  doi:10.1016/0550-3213(84)90461-9;
  H.~Goldberg,
  Phys.\ Rev.\ Lett.\  {\bf 50} (1983) 1419
   Erratum: [Phys.\ Rev.\ Lett.\  {\bf 103} (2009) 099905].
  doi:10.1103/PhysRevLett.50.1419
 
\bibitem{eft1}
  M.~Beltran, D.~Hooper, E.~W.~Kolb and Z.~C.~Krusberg,
  Phys.\ Rev.\ D {\bf 80} (2009) 043509
  doi:10.1103/PhysRevD.80.043509
  [arXiv:0808.3384 [hep-ph]].
  
\bibitem{eft2}
  M.~Beltran, D.~Hooper, E.~W.~Kolb, Z.~A.~C.~Krusberg and T.~M.~P.~Tait,
  JHEP {\bf 1009} (2010) 037
  doi:10.1007/JHEP09(2010)037
  [arXiv:1002.4137 [hep-ph]].

\bibitem{moreoneft}
  Y.~Bai, P.~J.~Fox and R.~Harnik,
  JHEP {\bf 1012} (2010) 048 \\
  doi:10.1007/JHEP12(2010)048
  [arXiv:1005.3797 [hep-ph]].

\bibitem{busoni}
  G.~Busoni, A.~De Simone, E.~Morgante and A.~Riotto,
  Phys.\ Lett.\ B {\bf 728} (2014) 412
  doi:10.1016/j.physletb.2013.11.069
  [arXiv:1307.2253 [hep-ph]].
  G.~Busoni, A.~De Simone, J.~Gramling, E.~Morgante and A.~Riotto,
  JCAP {\bf 1406} (2014) 060
  doi:10.1088/1475-7516/2014/06/060
  [arXiv:1402.1275 [hep-ph]].

 \bibitem{mccabe}
  O.~Buchmueller, M.~J.~Dolan and C.~McCabe,
  JHEP {\bf 1401} (2014) 025
  doi:10.1007/JHEP01(2014)025
  [arXiv:1308.6799 [hep-ph]].

\bibitem{WhitePapers}
 S.~A.~Malik {\it et al.}, {\it White Paper from 2014 Brainstorming Workshop held at Imperial College London},
  Phys.\ Dark Univ.\  {\bf 9-10} (2015) 51
  doi:10.1016/j.dark.2015.03.003
  [arXiv:1409.4075 [hep-ex]];
  J.~Abdallah {\it et al.},
  Phys.\ Dark Univ.\  {\bf 9-10} (2015) 8
  doi:10.1016/j.dark.2015.08.001
  [arXiv:1506.03116 [hep-ph]];
 M.~Bauer {\it et al.}, {\it White Paper from 2016 Brainstorming Workshop held at Imperial College London},
  arXiv:1607.06680 [hep-ex].

\bibitem{zportal}
  G.~Arcadi, Y.~Mambrini and F.~Richard,
  JCAP {\bf 1503} (2015) 018
  doi:10.1088/1475-7516/2015/03/018
  [arXiv:1411.2985 [hep-ph]].

\bibitem{higgs1}
  A.~Djouadi, O.~Lebedev, Y.~Mambrini and J.~Quevillon,
  Phys.\ Lett.\ B {\bf 709} (2012) 65
  doi:10.1016/j.physletb.2012.01.062
  [arXiv:1112.3299 [hep-ph]].

 \bibitem{higgs2}
  A.~Djouadi, A.~Falkowski, Y.~Mambrini and J.~Quevillon,
  Eur.\ Phys.\ J.\ C {\bf 73} (2013) no.6,  2455
  doi:10.1140/epjc/s10052-013-2455-1
  [arXiv:1205.3169 [hep-ph]].

\bibitem{Estonia}
J.~Ellis, A.~Fowlie, L.~Marzola and M.~Raidal,
  Phys.\ Rev.\ D {\bf 97} (2018) no.11,  115014
  doi:10.1103/PhysRevD.97.115014
  [arXiv:1711.09912 [hep-ph]].

\bibitem{langacker1984}
  P.~Langacker, R.~W.~Robinett and J.~L.~Rosner,
  Phys.\ Rev.\ D {\bf 30} (1984) 1470.
  doi:10.1103/PhysRevD.30.1470

  
\bibitem{erler}
  J.~Erler and P.~Langacker,
  Phys.\ Lett.\ B {\bf 456} (1999) 68
  doi:10.1016/S0370-2693(99)00457-8
  [hep-ph/9903476].

\bibitem{appelquist}
  T.~Appelquist, B.~A.~Dobrescu and A.~R.~Hopper,
  Phys.\ Rev.\ D {\bf 68} (2003) 035012
  doi:10.1103/PhysRevD.68.035012
  [hep-ph/0212073].
  
 \bibitem{Carena}
 M.~Carena, A.~Daleo, B.~A.~Dobrescu and T.~M.~P.~Tait,
  Phys.\ Rev.\ D {\bf 70} (2004) 093009
  doi:10.1103/PhysRevD.70.093009
  [hep-ph/0408098].


\bibitem{morrissey}
  D.~E.~Morrissey and J.~D.~Wells,
  Phys.\ Rev.\ D {\bf 74} (2006) 015008
  doi:10.1103/PhysRevD.74.015008
  [hep-ph/0512019].

\bibitem{chiang}
  C.~W.~Chiang, N.~G.~Deshpande and J.~Jiang,
  JHEP {\bf 0608} (2006) 075
  doi:10.1088/1126-6708/2006/08/075
  [hep-ph/0606122].

\bibitem{langacker}
  P.~Langacker,
  Rev.\ Mod.\ Phys.\  {\bf 81} (2009) 1199
  doi:10.1103/RevModPhys.81.1199
  [arXiv:0801.1345 [hep-ph]].
  

  
  
\bibitem{barr}
  S.~M.~Barr, B.~Bednarz and C.~Benesh,
  Phys.\ Rev.\ D {\bf 34} (1986) 235.
  doi:10.1103/PhysRevD.34.235


\bibitem{batra}
  P.~Batra, B.~A.~Dobrescu and D.~Spivak,
  J.\ Math.\ Phys.\  {\bf 47} (2006) 082301
  doi:10.1063/1.2222081
  [hep-ph/0510181].
  
\bibitem{ekstedt}  
  A.~Ekstedt, R.~Enberg, G.~Ingelman, J.~L喃gren and T.~Mandal,
  JHEP {\bf 1611} (2016) 071
  doi:10.1007/JHEP11(2016)071
  [arXiv:1605.04855 [hep-ph]].


 \bibitem{Ismail}
 A.~Ismail, W.~Y.~Keung, K.~H.~Tsao and J.~Unwin,
  Nucl.\ Phys.\ B {\bf 918} (2017) 220
  doi:10.1016/j.nuclphysb.2017.03.001
  [arXiv:1609.02188 [hep-ph]].

  
\bibitem{Kahlhoefer:2015bea}
  F.~Kahlhoefer, K.~Schmidt-Hoberg, T.~Schwetz and S.~Vogl,
  JHEP {\bf 1602} (2016) 016
  doi:10.1007/JHEP02(2016)016
  [arXiv:1510.02110 [hep-ph]].


\bibitem{Katz}
A.~Ismail, A.~Katz and D.~Racco,
  JHEP {\bf 1710} (2017) 165
  doi:10.1007/JHEP10(2017)165
  [arXiv:1707.00709 [hep-ph]].



\bibitem{usEFT1}
  J.~Ellis, M.~Fairbairn and P.~Tunney,
  JHEP {\bf 1708} (2017) 053
  doi:10.1007/JHEP08(2017)053
  [arXiv:1704.03850 [hep-ph]].

\bibitem{Duerr}
 P.~Fileviez Perez, S.~Ohmer and H.~H.~Patel,
  Phys.\ Lett.\ B {\bf 735} (2014) 283
  doi:10.1016/j.physletb.2014.06.057
  [arXiv:1403.8029 [hep-ph]];
 M.~Duerr, P.~Fileviez Perez and M.~B.~Wise,
  Phys.\ Rev.\ Lett.\  {\bf 110} (2013) 231801
  doi:10.1103/PhysRevLett.110.231801
  [arXiv:1304.0576 [hep-ph]];
 M.~Duerr and P.~Fileviez Perez,
  Phys.\ Rev.\ D {\bf 91} (2015) no.9,  095001
  doi:10.1103/PhysRevD.91.095001
  [arXiv:1409.8165 [hep-ph]].
 


  
 \bibitem{usEFT2}
  J.~Ellis, M.~Fairbairn and P.~Tunney,
  Eur.\ Phys.\ J.\ C {\bf 78} (2018) no.3,  238
  doi:10.1140/epjc/s10052-018-5725-0
  [arXiv:1705.03447 [hep-ph]].

\bibitem{tytgat}
  G.~Arcadi, Y.~Mambrini, M.~H.~G.~Tytgat and B.~Zaldivar,
  JHEP {\bf 1403} (2014) 134
  doi:10.1007/JHEP03(2014)134
  [arXiv:1401.0221 [hep-ph]].
  
\bibitem{XENON1T}
  E.~Aprile {\it et al.} [XENON Collaboration],
  arXiv:1805.12562 [astro-ph.CO].
  
  \bibitem{sandt2018}
  J.~Ellis, C.~W.~Murphy, V.~Sanz and T.~You,
  arXiv:1803.03252 [hep-ph].

 
 \bibitem{Yseq}
 T.~Appelquist, B.~A.~Dobrescu and A.~R.~Hopper,
  Phys.\ Rev.\ D {\bf 68} (2003) 035012
  doi:10.1103/PhysRevD.68.035012
  [hep-ph/0212073].
  
\bibitem{fairbairnheal}
  M.~Fairbairn and J.~Heal,
  Phys.\ Rev.\ D {\bf 90} (2014) no.11,  115019
  doi:10.1103/PhysRevD.90.115019
  [arXiv:1406.3288 [hep-ph]].

  


 
  
\bibitem{micromegas}
  G.~B{\' e}langer, F.~Boudjema, A.~Pukhov and A.~Semenov,
  Comput.\ Phys.\ Commun.\  {\bf 192} (2015) 322
  doi:10.1016/j.cpc.2015.03.003
  [arXiv:1407.6129 [hep-ph]].

\bibitem{feynrules}
  A.~Alloul, N.~D.~Christensen, C.~Degrande, C.~Duhr and B.~Fuks,
  Comput.\ Phys.\ Commun.\  {\bf 185} (2014) 2250
  doi:10.1016/j.cpc.2014.04.012
  [arXiv:1310.1921 [hep-ph]].

\bibitem{Chala:2015ama}
  M.~Chala, F.~Kahlhoefer, M.~McCullough, G.~Nardini and K.~Schmidt-Hoberg,
  JHEP {\bf 1507} (2015) 089
  doi:10.1007/JHEP07(2015)089
  [arXiv:1503.05916 [hep-ph]].
  
 \bibitem{AEMR}
 A.~Arbey, J.~Ellis, F.~Mahmoudi and G.~Robbins,
  arXiv:1807.00554 [hep-ph].
  
\bibitem{LHCZprime}
ATLAS Collaboration, {\tt https://cds.cern.ch/record/2259039/files/ATLAS-} {\tt CONF-2017-027.pdf};
see also CMS Collaboration, {\tt https://cds.cern.ch/record/} {\tt 2205764/files/EXO-16-031-pas.pdf};

\bibitem{madgraph}
  J.~Alwall {\it et al.},
  JHEP {\bf 1407} (2014) 079
  doi:10.1007/JHEP07(2014)079
  [arXiv:1405.0301 [hep-ph]].

 \bibitem{ddtools}
  M.~Cirelli, E.~Del Nobile and P.~Panci,
  JCAP {\bf 1310} (2013) 019
  doi:10.1088/1475-7516/2013/10/019
  [arXiv:1307.5955 [hep-ph]].
  
\bibitem{shoemaker}
  I.~M.~Shoemaker and L.~Vecchi,
  Phys.\ Rev.\ D {\bf 86} (2012) 015023
  doi:10.1103/PhysRevD.86.015023
  [arXiv:1112.5457 [hep-ph]].
 
\bibitem{fox}
  P.~J.~Fox, R.~Harnik, R.~Primulando and C.~T.~Yu,
  Phys.\ Rev.\ D {\bf 86} (2012) 015010
  doi:10.1103/PhysRevD.86.015010
  [arXiv:1203.1662 [hep-ph]].

\bibitem{BM}
O.~Buchmueller, S.~A.~Malik, C.~McCabe and B.~Penning,
  Phys.\ Rev.\ Lett.\  {\bf 115} (2015) no.18,  181802
  doi:10.1103/PhysRevLett.115.181802
  [arXiv:1505.07826 [hep-ph]].

\bibitem{babumix}
  K.~S.~Babu, C.~F.~Kolda and J.~March-Russell,
  Phys.\ Rev.\ D {\bf 57} (1998) 6788
  doi:10.1103/PhysRevD.57.6788
  [hep-ph/9710441].

\bibitem{duerrmix}
  M.~Duerr, F.~Kahlhoefer, K.~Schmidt-Hoberg, T.~Schwetz and S.~Vogl,
  JHEP {\bf 1609} (2016) 042
  doi:10.1007/JHEP09(2016)042
  [arXiv:1606.07609 [hep-ph]].

\bibitem{pdg}
  C.~Patrignani {\it et al.} [Particle Data Group],
  Chin.\ Phys.\ C {\bf 40} (2016) no.10,  100001.
  doi:10.1088/1674-1137/40/10/100001
  
\bibitem{Aaboud:2018zba}
  M.~Aaboud {\it et al.} [ATLAS Collaboration],
  arXiv:1801.08769 [hep-ex].

\bibitem{ATLAS:2016bvn}
  The ATLAS collaboration [ATLAS Collaboration],
  ATLAS-CONF-2016-070.

\bibitem{Aaboud:2018fzt}
  M.~Aaboud {\it et al.} [ATLAS Collaboration],
  arXiv:1804.03496 [hep-ex].

\bibitem{Aaboud:2017yvp}
  M.~Aaboud {\it et al.} [ATLAS Collaboration],
  Phys.\ Rev.\ D {\bf 96} (2017) no.5,  052004
  doi:10.1103/PhysRevD.96.052004
  [arXiv:1703.09127 [hep-ex]].
  
  \bibitem{dwarves}
  M.~L.~Ahnen {\it et al.} [MAGIC and Fermi-LAT Collaborations],
  JCAP {\bf 1602} (2016) no.02,  039
  doi:10.1088/1475-7516/2016/02/039
  [arXiv:1601.06590 [astro-ph.HE]].

\bibitem{Carone:1995pu}
  C.~D.~Carone and H.~Murayama,
  Phys.\ Rev.\ D {\bf 52} (1995) 484
  doi:10.1103/PhysRevD.52.484
  [hep-ph/9501220].

\bibitem{Jacques:2015zha}
  T.~Jacques and K.~Nordstr{\" o}m,
  JHEP {\bf 1506} (2015) 142
  doi:10.1007/JHEP06(2015)142
  [arXiv:1502.05721 [hep-ph]].
  
\bibitem{Aaboud:2017phn}
  M.~Aaboud {\it et al.} [ATLAS Collaboration],
  JHEP {\bf 1801} (2018) 126
  doi:10.1007/JHEP01(2018)126
  [arXiv:1711.03301 [hep-ex]].
  
  
  
  
  
  
  
  
  
  
  
  
  
  
  
  
  
  
  
  
  
  
  
  
  
  
  
\end{thebibliography}
\end{document}